\documentclass[sigconf]{acmart}

\def\BibTeX{{\rm B\kern-.05em{\sc i\kern-.025em b}\kern-.08em
    T\kern-.1667em\lower.7ex\hbox{E}\kern-.125emX}}

\usepackage{booktabs} % For formal tables
\usepackage{algorithm}
\usepackage{algpseudocode}
\usepackage{tikz}
\usetikzlibrary{trees}
\usepackage{amsmath}
\usepackage{url} 
\newtheorem{definition}{\sc Definition}[section]
\setcopyright{rightsretained}

\acmDOI{}

\acmISBN{978-3-98318-097-4}

\begin{document}
\title{\texttt{RIPOST}: Two-Phase Private Decomposition for Multidimensional Data}\titlenote{This work is funded by DigiTrust (\url{http://lue.univ-lorraine.fr/fr/article/digitrust/}).}

\author{Ala Eddine Laouir}

\orcid{0009-0002-1103-0312}
\affiliation{%
  \institution{Universit\'e de Lorraine, CNRS, Inria, LORIA}
  \city{F-54000 Nancy}
  \country{France}
}
\email{ala-eddine.laouir@loria.fr}

\author{Abdessamad Imine}

\orcid{0000-0002-4496-0340}
\affiliation{%
  \institution{Universit\'e de Lorraine, CNRS, Inria, LORIA}
  \city{F-54000 Nancy}
  \country{France}
}
\email{abdessamad.imine@loria.fr}

% The default list of authors is too long for headers}
% \renewcommand{\shortauthors}{B. Trovato et al.}
\renewcommand{\shortauthors}{}

\begin{abstract}
Differential privacy (DP) is considered as the gold standard for data privacy. While the problem of answering simple queries and functions under DP guarantees has been thoroughly addressed in recent years, the problem of releasing multidimensional data under DP remains challenging. 
In this paper, we focus on this problem, in particular on how to construct privacy-preserving views using a domain decomposition approach. 
The main idea is to recursively split the domain into sub-domains until a convergence condition is met. The resulting sub-domains are perturbed and then published in order to be used to answer arbitrary queries. 
Existing methods that have addressed this problem using domain decomposition face two main challenges: (i) efficient privacy budget management over a variable and undefined decomposition depth $h$; and (ii) defining an optimal data-dependent splitting strategy that minimizes the error in the sub-domains while ensuring the smallest possible decomposition.
To address these challenges, we present \texttt{RIPOST}, a multidimensional data decomposition algorithm that bypasses the constraint of predefined depth $h$ and applies a data-aware splitting strategy to optimize the quality of the decomposition results. 
The core of \texttt{RIPOST} is a two-phase strategy that separates non-empty sub-domains at an early stage from empty sub-domains by exploiting the properties of multidimensional datasets, and then decomposes the resulting sub-domains with minimal inaccuracies using the mean function. 
Moreover, \texttt{RIPOST} introduces a privacy budget distribution that allows decomposition without requiring prior computation of the depth $h$. 
Through extensive experiments, we demonstrated that \texttt{RIPOST} outperforms state-of-the-art methods in terms of data utility and accuracy on a variety of datasets and test cases\footnote{\url{https://github.com/AlaEddineLaouir/RIPOST.git}}.
\end{abstract}

%
% % The code below should be generated by the tool at
% % http://dl.acm.org/ccs.cfm
% % Please copy and paste the code instead of the example below. 
% %
% \begin{CCSXML}
% <ccs2012>
%  <concept>
%   <concept_id>10010520.10010553.10010562</concept_id>
%   <concept_desc>Computer systems organization~Embedded systems</concept_desc>
%   <concept_significance>500</concept_significance>
%  </concept>
%  <concept>
%   <concept_id>10010520.10010575.10010755</concept_id>
%   <concept_desc>Computer systems organization~Redundancy</concept_desc>
%   <concept_significance>300</concept_significance>
%  </concept>
%  <concept>
%   <concept_id>10010520.10010553.10010554</concept_id>
%   <concept_desc>Computer systems organization~Robotics</concept_desc>
%   <concept_significance>100</concept_significance>
%  </concept>
%  <concept>
%   <concept_id>10003033.10003083.10003095</concept_id>
%   <concept_desc>Networks~Network reliability</concept_desc>
%   <concept_significance>100</concept_significance>
%  </concept>
% </ccs2012>  
% \end{CCSXML}
% 
% \ccsdesc[500]{Computer systems organization~Embedded systems}
% \ccsdesc[300]{Computer systems organization~Redundancy}
% \ccsdesc{Computer systems organization~Robotics}
% \ccsdesc[100]{Networks~Network reliability}

% \keywords{ACM proceedings, \LaTeX, text tagging}

%% A "teaser" image appears between the author and affiliation
%% information and the body of the document, and typically spans the
%% page.

\maketitle

% PAPER STARTS

\sloppy

\section{Introduction}
Releasing  a privacy-preserving view of the data, which does not compromise the privacy of individuals whose information is stored in the dataset, has received significant attention from both the research community and major corporations\footnote{\url{https://www.nist.gov/ctl/pscr/open-innovation-prize-challenges/past-prize-challenges/2018-differential-privacy-synthetic}}. The differential privacy model has become the de facto standard for ensuring data privacy \cite{apple,google,census}. It guarantees that the presence or absence of any particular individual in the dataset does not substantially affect the outcome of the released view. This provides individuals with the plausible deniability needed to deny their inclusion in the data, thus ensuring their privacy.
The resulting privacy-preserving view can be used to answer OLAP (Online Analytical Processing) queries, create multi-dimensional histograms, and perform data mining tasks such as clustering \cite{privtree}. This versatility in the use of the released view has encouraged researchers to further explore this field.

A naive approach \cite{iden} would be to add noise to all data points, but this results in a significant loss of data utility. Therefore, more sophisticated mechanisms and techniques have been proposed in the literature. Some use generative models to produce synthetic differentially private data \cite{privbayes,dppro,gan1,gan2,gan3,p3gm,privsyn}, focusing primarily on image data or regular tabular data. Others \cite{mm,hdmm,dawa} propose workload-dependent algorithms that aim to release answers to a predefined batch of queries rather than the full data view. The generative-based solutions are generally designed as general-purpose algorithms (especially those utilizing deep learning models) without a specific focus on multidimensional data or tensors. On the other hand, workload-dependent approaches are constrained by their reliance on the query workload, which limits their usability. Moreover, these methods often have high computational complexity, making them less scalable for multidimensional data.

In this paper, we focus on decomposition-based algorithms, as they are well-suited for multidimensional data and provide useful views that can be leveraged for a variety of analytical tasks such as OLAP queries. In the literature, several differentially private domain decomposition methods have been proposed to release multidimensional data \cite{hdpview,privtree,ahp,cormode,depth1,depth2}. These solutions face two main challenges when constructing the view: 
(i) \textbf{The reliance on the decomposition depth $h$} to manage the privacy budget. Several methods \cite{cormode,depth1,depth2,ahp} estimate $h$ before starting the decomposition to limit its value. However, only \cite{hdpview,privtree} successfully overcome this limitation either completely or partially. 
(ii) \textbf{The use of data-aware domain splitting strategies} to ensure minimal error in sub-domain approximations while preserving the utility of the generated view. Notably, \cite{privtree} neglects this issue by applying a data-independent splitting strategy, while \cite{hdpview} applies data-aware splitting only partially, not during the entire decomposition process.
All of these domain decomposition methods fail to address both challenges successfully. In comparison with many existing methods, our proposed domain decomposition method, (\texttt{RIPOST} — p\textbf{R}ivate v\textbf{I}ew by two-\textbf{P}hase decomp\textbf{O}sition for multidimen\textbf{S}ional da\textbf{T}a), is able to fully overcome these limitations, producing a high-accuracy view compared to the others.

{\noindent\bf Our Contributions.}
Our proposed solution \texttt{RIPOST} eliminates the constraint of a predefined depth $h$, allowing for a more refined private decomposition as needed. 
This flexibility is achieved through a novel privacy budget distribution strategy that relies on a convergent series with a bounded sum as the allocation factor. 
The tree structure resulting from the decomposition of \texttt{RIPOST} is adapted to speed up query processing. 
However, the actual evaluation is performed at the leaf nodes, which are the constituents of the published private view \texttt{RIPOST} (similarly to existing works such as \cite{hdpview,privtree}). 

Regarding the decomposition strategy, \texttt{RIPOST} uses a novel data-aware splitting strategy that takes into account both the data distribution in the domain and the corresponding approximation error. 
\texttt{RIPOST} is a two-phase decomposition method that separates empty sub-domains from populated sub-domains, thus avoiding unnecessary decompositions and focusing only on highly populated sub-domains with large approximation errors. 
Those sub-domains are then passed to the second phase of decomposition to further refine them to minimize the error.

%\noindent\textbf{Preview of results.} 
Through extensive experiments, we compared our algorithm with various existing methods in the literature, evaluating their performance based on the error (measured as relative root mean squared error, R-RMSE) produced during query evaluation on the published view. Table \ref{tab:res_pre} presents the average Relative-RMSE obtained when comparing \texttt{RIPOST} with \cite{hdpview,privtree,p3gm,privbayes} using the Fire dataset\footnote{\url{https://data.sfgov.org/Public-Safety/Fire-Department-and-Emergency-Medical-Services-Dis/nuek-vuh3/data}}. Our approach achieved the smallest error on average, outperforming the closest competitor \cite{hdpview} by 47\% and showing more than a 10x improvement compared to other methods.\medskip

\begin{table}
    \centering
    \begin{tabular}{|c|c|c|c|c|c|} \hline 
         Algo&  \texttt{RIPOST}&  \cite{hdpview}&  \cite{privtree}&  \cite{p3gm}&  \cite{privbayes}\\ \hline 
         Avg R-RMSE&  1&  x1.47&  x13&  x22&  x62\\ \hline
    \end{tabular}
    \caption{Comparative results of \texttt{RIPOST}}
    \label{tab:res_pre}
\end{table}
% Our proposed method \texttt{RIPOST} offers an independece from for the contraint of the depth $h$, allowing for refined decomposition as needed. This flexibility is attributed to  a novel budget distribution strategy that uses convergent series with bounded sum as factore of allocation. The tree structure that results from the decomposition done \texttt{RIPOST}, is suitable for accelerating queries processing, but the actuel evaluation is done the leaf nodes. which are the constituents of the private view published by \texttt{RIPOST}.
% As for the decomposition strategy, \texttt{RIPOST} takes into account both the distribution of the data in the domain and their approximation error during the decomposition. It introduces a novel two-phase decomposition approach that aims to seperate the empty sub-domains from the populated sub-domains, and avoiding unnecessary decomposition, and focusing only on the sub-domains that are highly populated and have significant approximation errors. \medskip
% Additionally, \texttt{RIPOST} introduces a novel budget distribution strategy that enables recursive/consecutive  domain decomposition without any constraints on the depth, allowing the method to decompose as needed to improve the quality of the released view.\medskip

\noindent\textbf{Roadmap.} The paper is structured as follows: Section \ref{sec:rw} reviews related work. Section \ref{sec:pre} introduces the key concepts and notation used throughout the paper. Section \ref{sec:ps} describes the problem that \texttt{RIPOST} addresses. Section \ref{sec:ripost} presents the details of the \texttt{RIPOST} method. The extensive evaluation of the method is provided in Section \ref{sec:eval}. Section \ref{sec:dis} discusses potential limitations and extensions, and we conclude in Section \ref{sec:conclusion} by outlining key results and contributions.
\medskip

\section{Related Works}\label{sec:rw}
The problem of data publication has become a major research topic in recent years. The naive solution would be to inject noise into each cell of the \cite{iden,winslett} tensor, but this would cause too much perturbation, making the published data inaccurate. 
To provide a better alternative, many methods have been developed to publish and query high-dimensional data under DP. We present here the state-of-the-art on this problem.\medskip

\noindent{\bf Partitioning approaches.} 
Rather than perturbing data points, \cite{ahp,dawa} proposed algorithms that partition data into bins. In \cite{dawa}, they focused on data partitioning based on minimal \textit{aggregation error}, and then applied the \textit{Laplace mechanism} to add noise to the mean value of each bin. Their solution is only effective in low-dimensional/small-domain data, due to the complexity of finding the best partitioning. 
Other solutions have proposed a recursive domain decomposition algorithm that uses the quadtree strategy to divide the tensor into smaller blocks \cite{privtree,14}. 
Partitioning continues until a convergence condition is satisfied. After convergence, blocks will only keep a representative value of the cells, such as the mean (similar to \cite{dawa}), and perturb this value to ensure DP. 
One of the main challenges with partitioning is defining a limit or depth, denoted as $h$, to manage the budget distribution. Some solutions \cite{cormode,depth1,depth2} use heuristics to determine an appropriate value for $h$, while \cite{privtree} bypasses this constraint altogether, as it can hinder performance and introduce inaccuracies. We will address this challenge more in details in Section \ref{sec:ps}.

% One of the main challenges with partitioning is define a limit or depth $h$ in order to manage the budget distribution. Some solutions \cite{cormode,depth1,depth2}, use a heuristic to determine an $h$, meanwhile \cite{privtree} bypassed this strong constraint as it may hinder the performances and introduce inaccuracies.
HDPView \cite{hdpview} can be considered an improvement over \cite{privtree}. Instead of using quadtree partitioning, HDPView introduces a data-dependent partitioning strategy, which enhances the quality of the resulting DP view by reducing both the \textit{Aggregation Error (AE)} and DP noise. However, this data-dependent strategy is still constrained by the depth parameter $h$ (after which it becomes totally random). While the convergence of \cite{hdpview} is independent of $h$, the partitioning strategy itself is limited by $h$, which reintroduces the issue in part of their solution.

% HDPView \cite{hdpview} can be considered as an improvement of \cite{privtree}. Instead of using quadtree partitioning, they introduced a data-dependent partitioning strategy. Thus, the quality of the created DP view is improved in terms of the \textit{aggregation error} and the DP noise. But, this data-dependent strategy is constrained by the depth $h$. Which means that even if the convergence of \cite{hdpview} is independent of $h$, their partitioning strategy is, thus re-introduced the issue in part of their solution.
\medskip

\noindent{\bf Workload-based approaches.} 
Some solutions aim to release only a part of the data delimited by a predefined set of queries, also called \textit{Workload}. In \cite{mm}, they introduced a matrix mechanism (MM) that creates queries and outputs optimized for a given workload. The work in \cite{hdmm} presents an extension to \cite{mm} by adapting and optimizing the matrix mechanism to high-dimensional data (HDMM). 
In addition, HDMM improves \cite{dawa}, which also falls into this category of solutions. 
The main issues with these techniques are: (1) workload dependency, and (2) matrix operations can be computationally costly.
PrivateSQL \cite{privatesql} provides a framework that calculates the sensitivity of relational data based on its schema and relationships. It then creates a multidimensional view of these data using methods such as \cite{dawa,mm}.
% Another solution in this class is \cite{slimview}, which uses random sampling to create a synopsis of the original data and accompany it with a dedicated perturbation table to inject noise to each query. This solution is considered as a data publishing solution, beacuse it pertubs the data and no further budget is needed for the queries. But, it differs than the previosly introduced solution in a sense that the generated data remains at the server in order to manage the perturbation of the queries (can be considred as hybrid publishing/online query answering). Consequently, \cite{slimview} is relevent to our work but not quite comparable.
Another solution in this class is \cite{slimview}, which uses random sampling to create a synopsis of the original data and complements it with a dedicated perturbation table to inject noise into each query. This approach is considered a data publishing solution because it perturbs the data upfront, and no further budget is required for individual queries. However, it differs from the previously introduced solutions in that the generated data remains on the server to handle query perturbation, making it a hybrid between data publishing and online query answering. As a result, \cite{slimview} is relevant to our work but not directly comparable.
\medskip

\noindent{\bf Private generative models.} 
Another direction of research in this area is the use of generative models to create synthetic, private versions of the data that can then be published.  
In \cite{privbayes}, the authors trained a Bayesian network under DP guarantees, while others \cite{dppro,privsyn,priview} used marginal tables to learn the distribution of the original data.
Given the impressive performance of deep generative models in many research areas, \cite{gan,gan1,gan2,gan3,p3gm} proposed models that preserve DP while synthesizing data. 
These models, such as \cite{gan} (which uses a Generative Adversarial Network, GAN), outperform \cite{privbayes} only on some classification tasks, but \cite{privbayes} outperforms GAN-based methods on private tabular data. 
The limitations of these approaches are, in general, as follows: (1) The models are not specifically tailored to tensor data structures and OLAP tasks, which is a major drawback compared to partitioning and workload-dependent approaches, and (2) Generative models are very complex and resource-intensive for large-scale data. In scenarios where data is frequently updated, this computational cost becomes very significant.\medskip

In our work, we focus on \textbf{workload-independent partitioning-based approaches}. 
More specifically, we address the \textbf{limitations and inefficiencies} of existing methods such as \cite{hdpview,privtree,cormode} which we will present in more detail in Section \ref{sec:ps}. 
We propose a new partitioning algorithm \texttt{RIPOST}, which creates a better quality view in terms of \textbf{accuracy} while ensuring \textbf{data privacy and utility}.

%We propose a new partitioning algorithm called \texttt{RIPOST} (p\textbf{R}ivate v\textbf{I}ew by two-\textbf{P}hase decomp\textbf{O}sition for multidimen\textbf{S}ional da\textbf{T}a), which creates a better quality view in terms of \textbf{accuracy} while ensuring \textbf{data privacy and utility}.

\section{Preliminaries}\label{sec:pre}
In this section, we introduce the concepts and notation used throughout the paper.\medskip

\begin{figure}[!h]
    \centering
    \includegraphics[width=1\linewidth]{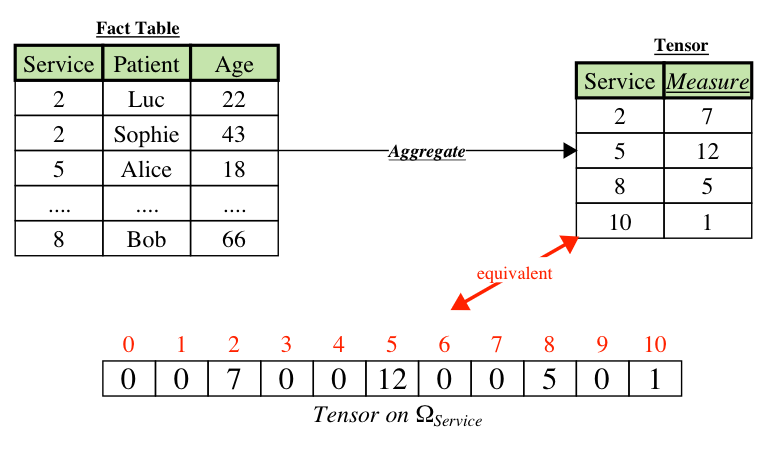}
    \caption{Data Model}
    \label{fig:dm}
\end{figure}

\noindent\textbf{Database.}
Let $\mathcal{B}$ be a database with a set of tuples defined over $|D|$ dimensions (or attributes) $D = \{d_1, \ldots, d_{|D|}\}$, where $|D|$ is called the \emph{number of dimensions} of $\mathcal{B}$. Each dimension $d \in D$ is associated with a domain (range) $\Omega_d = [d_s, d_e]$, containing $\vert \Omega_d \vert = d_e - d_s + 1$ discrete and totally ordered values. The overall domain size of $\mathcal{B}$ is $\vert \Omega_D \vert = \prod_{d \in D} \vert \Omega_d \vert$. 
In Figure \ref{fig:dm}, the database \textit{"Fact Table"} is shown with $D = \{Age, Service, Patient\}$ and $\Omega_{Service} = [0, 10]$\medskip

\noindent\textbf{Tensor.}
A tensor $\mathcal{T}$ is a \emph{multidimensional} data structure created by aggregating $\mathcal{B}$ based on a subset of $m\leq |D|$ dimensions in $D$. Such aggregation is done using a ``group-by'' query such as:

%\[
\noindent
$\texttt{SELECT } (d_1, \ldots, d_m, \texttt{aggregator}(.) \texttt{ AS measure})$ \\
$\texttt{ FROM B GROUP BY } (d_1, \ldots, d_m)$.
%\]

Here, $d_1, \ldots, d_m \in D$ are dimensions and $\texttt{aggregator}(.)$ is an aggregation function such as $\texttt{COUNT}$, $\texttt{SUM}$, $\texttt{AVG}$, etc., which takes an attribute (or dimension) as a parameter. 
Using $\texttt{COUNT}$ to build a tensor $\mathcal{T}$, the \texttt{measure} attribute contains the number of rows that have the same values (or coordinates) on the dimensions $d_1, \ldots, d_m$. 
In this case, $\mathcal{T}$ is called a \textit{count tensor}. In this work, we will focus on counting tensors\footnote{Note that our work can be generalized to any aggregation function}. For simplicity, we will refer to ``count tensor'' simply as ``tensor.''

In \emph{Fact Table} of Figure \ref{fig:dm}, the dimensions \emph{Patient} and \emph{Age} have been aggregated to create a tensor, which can be represented either as a \textit{multidimensional array} or as a \textit{tabular}. 
The first representation visualizes (and stores) the entire domain (e.g. tensor on $\Omega_{\texttt{Service}}$ in Figure \ref{fig:dm}), while the second stores only the non-empty($\neq 0$) cells of the domain. Since our goal is to protect the entire data (empty and non-empty cells), we therefore consider the entire domain.\medskip

\noindent\textbf{Count Range Queries.} We consider a special class of queries, namely \emph{count range queries}. A count range query $Q$ is defined on $k$ dimensions ($k \leq |D|$) with subranges $Q = \{\tau^Q_1 \subseteq \Omega_{d_1}, \ldots, \tau^Q_k \subseteq \Omega_{d_k}\}$, and it computes the sum of the cells within those ranges. Given the tensor in Figure \ref{fig:dm}, a query $Q$ defined by the following range: $Q = \{\texttt{Service} \in [7,10]\}$ will return: $6$. \medskip

\noindent\textbf{Differential Privacy (DP).} Differential Privacy is a privacy model that provides formal guarantees of \emph{indistinguishability}, ensuring that query results do not reveal significant information about the presence or absence of any particular individual. Consequently, it hides information about which of the \textit{neighbouring tensors} \cite{dp} was used to answer the query.

\begin{definition}[Neighbouring Tensors]
Two tensors $T$ and $T'$ are \emph{neighbouring} if we can obtain one of them by incrementing the value of a cell in the other by 1. We denote this distance as $Dist(T, T') = 1$.
\end{definition}

Let $M$ be a randomized data publishing algorithm. $M$ is \emph{differentially private}, when it is insensitive to the presence or absence of any individual in $T$.

\begin{definition}[$\epsilon$-Differential privacy  \cite{dppre}]
An algorithm $M$ satisfies $\epsilon$-\emph{differential privacy} (or $\epsilon$-\emph{DP}) if for any neighbouring tensors $T$ and $T'$, and all possible sets of outputs $S$: 
$$\frac{\mbox{Pr}\left[ M\left( T \right) \in  S \right]}{\mbox{Pr}\left[ M\left( T' \right) \in  S \right]} \leq e^\epsilon$$
The parameter $\epsilon$ is called the \emph{privacy budget}.
\end{definition}

It is well known that DP is used to answer specific queries on databases. Let $f$ be  a query on a tensor $\mathcal{T}$ whose its answer $f(\mathcal{T})$ returns a number. The \emph{sensitivity} of $f$ is the amount by which the output of $f$ changes for all neighbouring tensors.

\begin{definition}[Sensitivity \cite{dppre}]
The sensitivity of a query $f$ for
any two neighboring tensors $\mathcal{T}$, $\mathcal{T}'$ is:
$$\Delta_f = \max_{\mathcal{T},\mathcal{T}'} \left\| f(\mathcal{T}) - f(\mathcal{T}') \right\|_1$$
where $\left\| . \right\|_1$ is the $L_1$ norm.
\end{definition}

For instance, if $f$ is a count range query then $\Delta_f$ is $1$.

To achieve DP for a query $f$, we need  to add random noise to its answer $f(\mathcal{T})$. To inject adequate noise, some fundamental mechanisms have been developed for DP. One of them is the \emph{Laplace Mechanism}.

\begin{definition}[Laplace Mechanism \cite{dppre}]\label{def:lap}
The \emph{Laplace mechanism} adds noise to $f(\mathcal{T})$ as:
$$ f(\mathcal{T})+ Lap\left( \frac{\Delta_f}{\epsilon}\right)$$

where $\Delta_f$ is the \emph{sensitivity} of $f$, and $Lap(\alpha)$ denotes sampling from the Laplace distribution with center $0$ and scale $\alpha$.
\end{definition}

For instance, if $f$ is a count range query then the noise is sampled from the distribution $Lap(1/\epsilon)$.

Unlike the Laplace Mechanism (based on noisy numerical answers), the \emph{Exponential mechanism} allows for selecting the “best” element (according to a scoring function) in a set while preserving DP \cite{dppre}.

\begin{definition}[Exponential Mechanism \cite{dppre}]\label{def:em}
Given a set of values $V$ extracted from a tensor $\mathcal{T}$ and a real scoring function $u$ to select values in $V$. The \emph{Exponential Mechanism} randomly samples $v$ from $V$ with probability proportional to:
$$\exp\left( \frac{\epsilon \times u(\mathcal{T},v)}{2\times \Delta_u} \right)$$
where $\Delta_u$ is the sensitivity of $u$.
\end{definition}

The Exponential Mechanism satisfies DP by approximately maximizing the score of the value it returns, while sometimes returning a value from the set that does not have the highest score.

Combining several DP mechanisms is possible, and the privacy accounting is managed using the sequential and the parallel composition properties of DP. Let $M_1,...,M_n$ be mechanisms satisfying $\epsilon_1,...,\epsilon_n$ -DP.

\begin{theorem}[Sequential Composition  \cite  {dppre}]\label{th:seq}
An algorithm sequentially applying $M_1,...,M_n$ satisfies $\left( \sum_{j = 1}^{n}\epsilon_j \right)$-DP.
\end{theorem}

\begin{theorem}[Parallel Composition  \cite  {dppre}]\label{th:paral}
An algorithm applying $M_1,\ldots,M_n$ to $n$ disjoint datasets
%the disjoint datasets $B_1,...,B_n$ 
in parallel satisfies ($max^{n}_{j=1} \epsilon_j$)-DP.
%($\max_{j \in \left[ n \right]}\epsilon_j $)-DP.
\end{theorem}

\section{Problem Statement}\label{sec:ps}
% Given a multi-dimensional tensor $T$,  the spatial decomposition problem consist of splitting the domain and cells of  $T$  into disjoint sub-blocks (sub-domains) such as the cells within each block can be replaced with the their mean with minimal error. This error is called \textit{Aggregation Error or AE}, and computes how much the cells deviates from the being perfectly uniform. Range queries are then estimated on these sub-blocks, based on their intersection with each block. Hence, the minimization of the \textbf{AE} is very crucial for the utility. 
% Another layer of complexity is added when the decomposition is made \textit{DP}, since each block must be noised introducing another error called perturbation error (PE).  The relation between these two errors is: the more blocks generated, the less AE there is and the more PE there is, and vice versa. So the perfect decomposition reach a balance between the AE and PE.

Given a multidimensional tensor $T$, the spatial decomposition problem involves dividing the domain and cells of $T$ into disjoint sub-blocks (sub-domains) such that the cells within each block $b$ can be approximated by their mean $\bar{x}$ with minimal error. This error, referred to as \textit{Aggregation Error} (\textbf{AE}), measures how much the cells deviate from being perfectly uniform $AE = \sum_{c\in b} \vert \bar{x} - c \vert$, where $c$ is a cell in $b$. Range queries are estimated on these sub-blocks based on their intersections with the blocks. Consequently, minimizing \textbf{AE} is essential for maximizing utility.

% In order to create such sub-domains (blocks), a decomposition algorithm two main components: 
% \begin{itemize}
%     \item A convergence condition(CC), which indicates whether further partitioning is needed or not.
%     \item A splitting strategy(SS) to divide a given domain into multiple smaller domains.
% \end{itemize}
% So the decomposition process  creates  a tree like structure that could be used as as index to speed up the access to blocks with in a certain domain.
In order to create such sub-domains (blocks), a decomposition algorithm consists of two main components:  
\begin{itemize}  
    \item A convergence condition (\textbf{CC}), which determines whether further partitioning of a domain is necessary.  
    \item A splitting strategy (\textbf{SS}), which defines how a given domain (sub-domain) is divided into smaller sub-domains.  
\end{itemize}  
As a result, the decomposition process produces a tree-like structure that serves as an index, enabling faster access to blocks within a given domain.

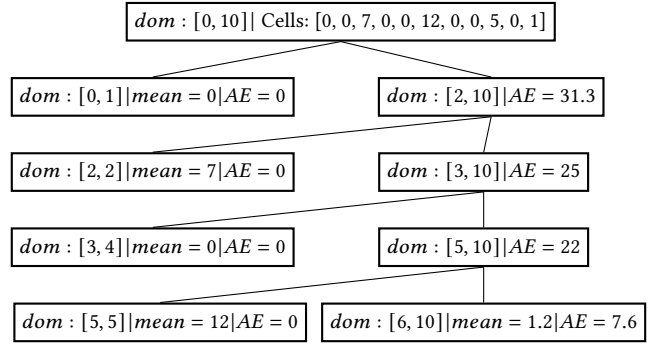
\begin{figure}[H]
    \centering
   \begin{tikzpicture}[
  grow via three points={one child at (0,-.8) and two children at (-3,-1) and (3,-1)},
  edge from parent path={(\tikzparentnode.south) -- (\tikzchildnode.north)},
  level distance=3cm,
  sibling distance=8cm,
  every node/.style={draw, font=\footnotesize, text centered, anchor=north},
  array/.style={draw, minimum height=.5cm, minimum width=1.5cm, anchor=center, font=\small}
]

% Root array (Full array)
\node (root) [array,thick] { $ dom:[0,10]|$ Cells: [0, 0, 7, 0, 0, 12, 0, 0, 5, 0, 1]}
  % Level 1: First partition
  child [xshift=.5cm]   { node (p1) [array,thick] {$ dom:[0,1] |mean=0 | AE=0$ } 
  }
  child [xshift=-1cm]  { node (p2) [array,thick] {$ dom:[2,10] | AE=31.3$}
    % Level 2: Split P2 further
    child [xshift=-1.5cm]   { node (p21) [array,thick] {$ dom:[2,2]|mean=7 | AE=0$} }
    child [xshift=-3.1cm]  { node (p22) [array ,thick] {$dom:[3,10] | AE=25$}
        child [xshift=-1.4cm]  { node (p221) [array,thick] {$dom:[3,4] |mean=0 | AE=0$}}
        child [xshift=-3cm] { node (p222) [array,thick] {  $ dom:[5,10] | AE=22$}
            child [xshift=-1.3cm]  { node (p2221) [array,thick] {$dom:[5,5] |mean=12 | AE=0$}}
            child [xshift=-3cm]  { node (p2222) [array,thick] { $dom:[6,10] |mean=1.2 | AE=7.6$}}
        }
    }
  };

\end{tikzpicture}
    \caption{Example of a decomposition Tree}
    \label{fig:exp_tree}
\end{figure}

% In figure \ref{fig:tree},  we see an example of decomposition process that splits the a domain as long as their is significant number of non empty cells.
In Figure \ref{fig:exp_tree}, an example of the tree created by the decomposition algorithm of \cite{hdpview} (without DP) is shown, where:

\begin{itemize}
    \item \textbf{CC} is defined based on the \textit{Aggregation Error} of the current sub-domain: $AE \le \theta$, where in this example, $\theta = 10$.
    \item \textbf{SS} splits a domain into two sub-domains, $\{ \text{left}, \text{right} \}$, and chooses the splitting point that minimizes $ \{ AE(\text{left}) + AE(\text{right}) \} $.
\end{itemize}

The illustrative example in Figure \ref{fig:exp_tree}, starts with one dimensional tensor with domaine $[0,10]$. Using \textbf{SS}, the process continue to split the domain and cells until the \textbf{CC} is satisfied. Notice that the mean is added only for the leaf nodes in Figure \ref{fig:exp_tree}. 
For approximating queries, only the leaves are considered. For example, for a query \( Q = \{\texttt{Service} \in [7,10]\} \), we need to compute its intersection with each leaf node \( ln \) in the list of tree leaves \( TL \) (\( ln \in TL \)). The approximation is therefor performed as follows:

\[
A(Q) = \sum_{ln \in TL} \left( ln_{\text{dom}} \cap Q_{\text{dom}} \right) \times ln_{\text{mean}}
\]

For this example query, the resulting approximation is \(A(Q) =  4.8 \).

% Notice that, only for leaf nodes we added the mean.
% As  for approximating queries, only the leaves are used. For instance, for a query $Q = \{\texttt{Service} \in [7,10]\}$ we need to compute its intersection with each leaf node $ln $ in list of  tree leaves $TL$, ($ln \in TL$), and the approximation is done as follows: $A(Q) = \sum_{ln\in TL} (ln_{dom}  \cup Q_{dom}) \times ln\_{mean}$. For this example query, the answer is $4.8$.

% The first challenge faced when creating a DP decomposition algorithm is to determine  how many times the CC should be tested? Given that CC is based on the data,  each test will need to consume a budget.  Since each branch operates on disjoint portion of the domain, to limit and distribute the budget the maximum depth of the tree must known in advance. 
The problem becomes more complex when the decomposition is made \textit{differentially private (DP)}, as each block must be noised, introducing an additional error termed \textit{Perturbation Error} (\textbf{PE}). These two errors are inversely related: increasing the number of blocks reduces \textbf{AE} but increases \textbf{PE}, and vice versa. Therefore, an optimal decomposition strikes a balance between \textbf{AE} and \textbf{PE}.
The first challenge in creating a DP decomposition algorithm is determining how many times the convergence condition (\textbf{CC}) should be tested. Since \textbf{CC} is data-dependent, each test consumes a portion of the privacy budget. Moreover, because each branch operates on a disjoint portion of the domain, only the maximum depth of the tree must be predefined to limit and appropriately distribute the budget.

% Defining this maximum depth $h$ independently of the underline data \cite{cormode, ali}, may facilitate the budget management but does not necessarily ensures an acceptable decomposition in regards of AE and PE. Using a heuristic prior to decomposition just to determine an adequate $h$ based on the data is not reasonable for two reasons: the result of the heuristic must be DP since it is based on the data, and secondly the complexity of finding $h$ may be similar to doing the actual decomposition.
Defining this maximum depth $h$ independently of the underlying data \cite{cormode,inan} can simplify budget management but does not necessarily guarantee an acceptable decomposition with respect to \textbf{AE} and \textbf{PE}. On the other hand, using a heuristic \cite{depth1,depth2} before decomposition to determine an appropriate $h$ based on the data is not practical for two reasons: (1) the result of the heuristic must be DP, as it relies on the data, and (2) the computational complexity of finding optimal $h$ may be comparable to performing the actual decomposition itself, thus they usually approximate $h$ vaguely.

% The first to bypass this dependency on $h$ is \cite{privtree}, by introducing a \textit{Baised convergence condition (BCC)} that can be repeated as much as needed while guaranteeing the budget consumption if fixed and independent of $h$. The BCC enforces a deterministic and an incrementing perturbation to the measure used in CC in addition to the \textit{Laplace noise}. This deterministic perturbation, referred to as the \textit{ bias term} in \cite{privtree},  is defined independently of the data and it may lead to what we call \textit{Early Convergence}. Where the decomposition stops before reaching an acceptable balance between AE and PE. 

The first approach to bypass the dependency on $h$ is PrivTree \cite{privtree}, which proposed a \textit{Biased Convergence Condition (BCC)}. This condition allows repeated testing while ensuring that the total privacy budget consumption remains fixed and independent of $h$. To better clarify the \textit{BCC}, consider a \textbf{CC} defined as $M(t) \le \theta$, where $M$ is any arbitrary metric, $t$ is a sub-block, and $\theta$ is a threshold. To ensure a DP-safe \textbf{CC}, \textit{Laplace} noise with a budget $\epsilon$ is added to $M(t)$, resulting in $M'(t)$, which is used for testing.  Their contribution relies on a key observation regarding budget consumption, particularly when:
\begin{itemize}
    \item $M(t) \gg \theta$ for all intermediate nodes in a path except the leaf, and
    \item For any parent/child pair $t_p$ and $t_c$ in a path, $M(t_p) - M(t_c) \ge c$.
\end{itemize}
Under these conditions, the actual budget consumption along the path is less than $ h / \epsilon$ (without releasing the actual value of $M'(t)$). They proved that if these two conditions are met, the total budget is constant and equal to $\epsilon$.

On real datasets, the issue arises with the second condition, as it cannot be guaranteed for all possible datasets. To address this, they introduced the \textit{BCC}, defined as follows:
\begin{itemize}
    \item $\hat{M(t)} = \max \{ \theta - \delta, M(t) - \text{depth}(t) \times \delta \}$, where $\delta$ is the \textit{bias} term, defined independently of the data. This guarantees that the second condition always holds: $\hat{M(t_p)} - \hat{M(t_c)} \ge \delta$. Alternatively, $\hat{M(t)} = \theta - \delta$, which is data-independent and does not require privacy budget.
    \item $M'(t) = \hat{M(t)} + \text{Lap}(\lambda)$, where $\lambda \ge \Phi \cdot \frac{1}{\epsilon}$, where of $\Phi$ is based on some parameters (see \cite{privtree} for more details). This ensures that the total budget of \textbf{CC} is $\epsilon$, regardless of how many times it is repeated.
\end{itemize}

This proposition indeed eliminates the need to specify a fixed depth at the beginning, but it has a significant disadvantage. The \textit{BCC} incorporates a deterministic, incrementing perturbation ($- \text{depth}(t) \times \delta$) to the metric $M$ used in \textbf{CC}, in addition to the standard \textit{Laplace noise}. This deterministic perturbation, referred to as the \textit{bias term} in \cite{privtree}, is defined independently of the data. Consequently, it may lead to what we call \textit{Early Convergence}, where the decomposition halts before achieving an acceptable balance between \textbf{AE} and \textbf{PE}.

% This proposiition indeed eliminates the need to specify a fixed depth at the beginng, but it have a significant disadvantage. The \textit{BCC} incorporates a deterministic, incrementing perturbation ($- depth(t)\times \delta$) to the measure used in the convergence condition (\textbf{CC}), in addition to the standard \textit{Laplace noise}. This deterministic perturbation, referred to as the \textit{bias term} in \cite{privtree}, is defined independently of the data. Thus, it may lead to what we call \textit{Early Convergence}, where the decomposition halts before achieving an acceptable balance between \textbf{AE} and \textbf{PE}.

\begin{figure}
    \centering
    \includegraphics[width=1\linewidth]{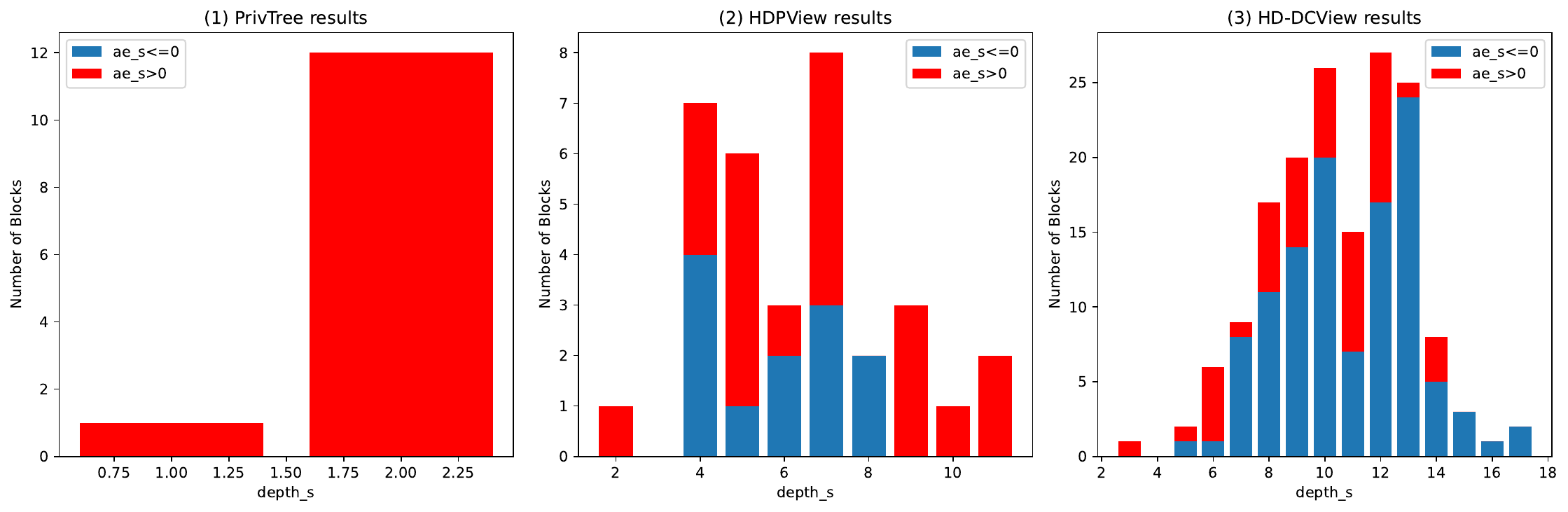}
    \caption{Convergence Analysis}
    \label{fig:conv_a}
\end{figure}

To assert such a claim, we conducted an experiment where we passed a 2D-tensor created using the dataset Adult\footnote{\url{https://archive.ics.uci.edu/dataset/2/adult}} to both PrivTree \cite{privtree} and HDPView \cite{hdpview}, since they both implement the \textit{BCC}. We set the total budget to $\epsilon = 0.1$. The results are shown in Figure \ref{fig:conv_a}, where plots 1 and 2 correspond to PrivTree and HDPView, respectively. The plots show how many sub-domains converged at each depth. Additionally, the red portion indicates the number of sub-domains with $AE > 0$, while the blue portion represents those with $AE = 0$.

% To assert such a claim, we conduct  an experiement, where we passsed a 2d-tensor created using dataset Adult \footnote{\url{https://archive.ics.uci.edu/dataset/2/adult}} to both PrivTree\cite{privtree} and HDPView \cite{hdpview} since they the \textit{BCC}. We set the total budget to $\epsilon = .1$. The results are shown in Figure \ref{fig:conv_a}, in which plots 1 and 2 related PrivTree and HDPView respectely, shows how many sub-domains has converged at each level. Also the portion in red the number of sub-domain with $AE > 0$ and in bleu the portion with $AE = 0$.

% We notive from the results, that HDPView generated more blocks thus having more blocks with $AE =0$. This highlight another issue with the \textit{BCC}, is that the choice of the metric $M$ is very important. Even thougth PrivTree uses the count as metric which has a \textit{sensitivity} less than the $AE$,  the deterministic and incrementing perturbation of the \textit{BCC} affected it greatly. To summerize the results in terms of early convergence, PrivTree hase $92\%$ of the subdomains with $AE$ and HDPView has $63\%$. In both cases these proportion are very significant, indicating the limit of the \textit{BCC}.

We notice from the results that HDPView generated more blocks, resulting in more blocks with $AE = 0$. This highlights another issue with the \textit{BCC}: the choice of the metric $M$ is very important. Even though PrivTree uses the count as a metric, which has a \textit{sensitivity} lower than that of $AE$ (see Section \ref{sec:sensi}), the deterministic and incrementing perturbation of the \textit{BCC} affected it greatly. 

To summarize the results in terms of \textit{Early Convergence}, PrivTree has $92\%$ of the sub-domains with $AE > 0$, while HDPView has $63\%$. In both cases, these proportions are very significant, indicating the limitations of the \textit{BCC}.

% The second challenge is the quality of the SS step of the decomposition algorithm. The SS can be defined as data independent process, for instance: at each level split the domain into $2^d$  \cite{privtree}.  Such strategy leads to creating to many unnecessary blocks that will increase substantially the PE with out as significant reduction in AE. On the other hand, using a data dependent SS  means that the a DP mechanism must be used and a budget to be consumed at each level, thus introducing the problem $h$ in the SS \cite{hdpview}. Also, the intuitive solution of designing an SS that uses AE as metric to split does not guarantee the best result since it ignores other defining characteristics of tensors that may help to improve the SS such as the sparsity.

The second challenge lies in the quality of the splitting strategy (\textbf{SS}) used in the decomposition algorithm. The \textbf{SS} can be defined as an data independent process; for example, splitting the domain into $2^{\vert D \vert}$ blocks at each level, as proposed in \cite{privtree}. However, such a strategy often creates an excessive number of unnecessary blocks, significantly increasing the \textbf{PE} without meaningfully reducing the \textbf{AE}. Conversely, employing a data-dependent \textbf{SS} requires a DP mechanism, which consumes privacy budget at each level. This reintroduces the dependency on $h$ within the \textbf{SS}, like in \cite{hdpview}. Furthermore, an intuitive solution, such as designing an \textbf{SS} that uses \textbf{AE} as the primary metric for splitting, does not necessarily yield the best results. This approach neglects other important characteristics of tensors, such as sparsity, which could be leveraged to improve the quality of the splitting strategy.

% With our proposed solution, we aim to address both challenges while ensuring to bypass all the limitation of previous solution. Given that all the multidimensional tensor are very spares, in a first phase our decomposition algorithm starts by separating the empty and the non empty regions in the tensor. In a second phase, it further decompose non-empty regions based on AE.   This separation allows us to minimise the AE with much less steps, because as lon as their a mixture of cells AE will no be null.  Additionally, our algorithm ensures that the CC and the SS in each phase are independent of the depth $h$ and does not require a fixed or maximum $h$ as input.
With our proposed solution \texttt{RIPOST}, we aim to address both challenges while overcoming the limitations of previous approaches. Leveraging the fact that multidimensional tensors are typically sparse, \texttt{RIPOST} operates in two phases. In the first phase, it separates the empty regions from the nonempty regions of the tensor. In the second phase, the nonempty regions are further decomposed based on \textbf{AE}.
% This two-phase separation allows us to minimize the \textbf{AE} in significantly fewer steps, as \textbf{AE} cannot be null as long as there is a mixture of cells. The tree example shown in Figure \ref{fig:exp_tree} enforces this logique, as seen in the results in Figure \ref{fig:exp_tree}  even thought the \textbf{SS} is based en \textbf{AE} it end-up seperating the empty and the non-empty cell. Furthermore, our algorithm ensures that both the convergence condition (\textbf{CC}) and the splitting strategy (\textbf{SS}) in each phase are independent of the depth $h$ and do not require a predefined or maximum $h$ as input.
Since  \textbf{AE} cannot be null as long as there is a mixture of cells, this two-phase \textbf{SS} allows us to minimize the \textbf{AE} in significantly fewer steps. The tree example shown in Figure \ref{fig:exp_tree} reinforces this logic. As seen in the results in Figure \ref{fig:exp_tree}, even though the \textbf{SS} is based on \textbf{AE}, it ends up separating the empty and non-empty cells. 

Furthermore, \texttt{RIPOST} ensures that both  \textbf{CC} and  \textbf{SS} in each phase are independent of the depth $h$ and do not require a predefined or maximum $h$ as input. In Figure\ref{fig:conv_a} plot 3, we see the performances of \texttt{RIPOST} compared to the others. It is able to reduce the \textit{Early Convergence} by a big margin to only $29\%$ while generating more refined blocks.

\section{Proposed Algorithm}\label{sec:ripost}
In this section, we present our method \texttt{RIPOST}, which constructs a DP view by privately decomposing a tensor into a set of blocks, and the user can query this view without any privacy risk while ensuring the utility of the query results.
\subsection{Overview of \texttt{RIPOST}}
Our proposed solution addresses all the key challenges and problems highlighted in Section \ref{sec:ps}. Thus, \texttt{RIPOST} iteratively splits the original tensor into two sub-blocks and stops when the cells in a block have minimal \textbf{AE}. To achieve such a decomposition, our algorithm operates in two phases: first, it tries to separate the empty cells from nonempty cells; then, it minimizes the \textbf{AE} in each block. 

During the first phase, the algorithm uses the $Sum$ of cells to determine whether it has successfully separated the empty cells ($Sum \leq 0$). This condition will be satisfied for blocks with nonempty cells due to perturbation. In the second phase, all blocks that converged based on the first condition are tested again and split to minimize the \textbf{AE}.
The overall algorithm of \texttt{RIPOST} is described in Algorithm \ref{alg:over}. \texttt{RIPOST} works as follows: 
\begin{itemize}
    \item As input, it takes a tensor $T$, a total budget $\epsilon$ and hyperparameters $\alpha,\gamma,\beta$ which are used to divide the budget over the first and second phases (Lines 1-4). 
    % PArticularly, $\alpha$ splits $\epsilon$ between the decomposition $\epsilon_d$ and perturbation $\epsilon_p$. $\gamma$ splits $\epsilon_d$ on the two phases$\epsilon_1, \epsilon_2$ of \texttt{RIPOST}, and $\beta$ splits $\epsilon_1$ and $\epsilon_2$ between convergence and splitting.
    \item The first phase checks $sum \leq \theta_1$ in the current block (Line 9). In case $Secure\_cc$ is \textbf{False}, the current block is still in the first phase. Then, it is divided into two sub-blocks by choosing the optimal cutting point that separates the cells and creates sub-blocks with a minimal mix of non-empty and empty cells (Line 12).%This splitting strategy ensures that the empty portions are detected/created early on, which helps early convergence of a particular portion of blocks with minimal splitting.
    \item In case $Secure\_cc$ is \textbf{True}, the current block is added to the list of $\textbf{converged\_blocks}$ (Line 10).
    \item This process is repeated until there are no more $\textbf{non\_converged\_blocks}$ (Line 7). 
    
    \item In the second phase, all the converged blocks from the previous phase are now considered as non-converged (Lines 17-18). 
    \item Each block with a minimal \textbf{AE} is added to the list of $\textbf{converged\_blocks}$ (Line 22). Otherwise, it is divided into two sub-blocks by choosing the optimal cutting point that minimizes \textbf{AE} in both sub-blocks (Line 24).
    \item This process is repeated as long as $\textbf{non\_converged\_blocks}$ (Algorithm \ref{alg:over} Line 19) is not empty.
  
  \item At the end, the average (representative value) of the $\text{converged\_blocks}$ is perturbed and the 
  $\textbf{secure\_view}$ is thus returned (Line 29). 
\end{itemize}

\begin{algorithm}[!h]
\caption{Our Partitioning algorithm \texttt{RIPOST}}\label{alg:over}
\begin{algorithmic}[1]
\Require $T$: Tensor; $\epsilon$:budget; $\alpha,\gamma,\beta$: hyperparameters
\State $\epsilon_p,\epsilon_d \gets \epsilon \times (1- \alpha),\epsilon \times \alpha $
\State $\epsilon_1,\epsilon_2 \gets \epsilon _d \times \gamma, \epsilon_d  \times (1-\gamma)  $
\State $\epsilon_{cc}^1,\epsilon_{ss}^1 \gets \epsilon_1 \times \beta, \epsilon_1 \times (1 - \beta)$
\State $\epsilon_{cc}^2,\epsilon_{ss}^2 \gets \epsilon_2 \times \beta, \epsilon_2 \times (1 - \beta)$

\State $\text{converged\_blocks} \gets []$
\State{$\text{non\_converged\_blocks} \gets [T]$}

\While{$\text{non\_converged\_blocks} \neq \emptyset $}  \Comment{First phase}
    \State $block \gets \text{non\_converged\_blocks.pop()} $
    \If{$Secure\_cc(block,Sum, \leq, \theta_1, \epsilon_{cc}^1)$} 
    \State $converged\_blocks.add(block)$
           
\Else   
     \State{$left\_block,right\_block \gets Secure\_ss(Sum,block,\epsilon_{ss}^1)$}
     \State{$\text{non\_converged\_blocks.push(left\_block)}$}
     \State{$\text{non\_converged\_blocks.push(right\_block)}$}
\EndIf 
\EndWhile

\State{$\text{non\_converged\_blocks} \gets converged\_blocks$}
\State $\text{converged\_blocks} \gets []$

\While{$non\_converged\_blocks \neq \emptyset$}  \Comment{ second phase}
\State $block \gets non\_converged\_blocks.pop()$
 \If{$Secure\_cc(block,AE, \leq, \theta_2, \epsilon_{cc}^2)$}
                    \State{$\text{converged\_blocks.push(block)}$}
            \Else  
                    \State{$left\_block,right\_block \gets Secure\_ss(AE,block,\epsilon_{ss}^2)$}
                    \State{$\text{non\_converged\_blocks.push(left\_block)}$}
                    \State{$\text{non\_converged\_blocks.push(right\_block)}$}
            \EndIf
\EndWhile
\State $secure\_view  \gets perturb(converged\_blocks)$
\State $\textbf{return secure\_view} $
\end{algorithmic}
\end{algorithm}

Regarding the budget distribution (Lines 1-4): \( \alpha \) splits the total privacy budget \( \epsilon \) between the decomposition budget \( \epsilon_d \) and the perturbation budget \( \epsilon_p \). The parameter \( \gamma \) then splits \( \epsilon_d \) over both phases of \texttt{RIPOST}, namely \( \epsilon_1 \) and \( \epsilon_2 \). Lastly, \( \beta \) divides \( \epsilon_1 \) and \( \epsilon_2 \) between the convergence and splitting components.

Note that we pass as parameter the entire budget allocated to the $Secure\_cc$ (resp. $Secure\_ss$) function in the current phase (Lines 9, 12, 21, 24). Since these operations are repeated many times, each iteration should only consume a part of the budget.
In Section \ref{sec:bd}, we will explain how to distribute the privacy budget over an unlimited number of iterations without exceeding the allocation. Then, in Section \ref{sec:cc_ss}, we will provide a detailed implementation of the $Secure\_cc$ and $Secure\_ss$ functions.

\subsection{Privacy Budget Distribution}\label{sec:bd}
In order to ensure that our algorithm \texttt{RIPOST} satisfies $\epsilon$-DP, we need to ensure not only that randomness is introduced, but also that the total budget remains below $\epsilon$ as required.
As mentioned in Section \ref{sec:ps}, controlling the budget in a decomposition often results in a limitation on depth, which introduces significant challenges and tradeoffs.
To avoid such a constraint, we need to define a function that distributes a given budget $\epsilon$ for any arbitrary depth $h$, such that:
\begin{equation}\label{eq:ec1}
    \epsilon \ge \sum_{i=1}^h{\epsilon_i} \text{, where } \epsilon_i > 0
\end{equation}
Since each $\epsilon_i < \epsilon$, we can define $\epsilon_i$ as: $\epsilon_i = \epsilon \times \omega_i$, where $\omega_i \in ]0,1[$. Thus, we can redefine the constraint in Equation \ref{eq:ec1} as:

\begin{equation}\label{eq:ec2}
    \begin{array}{ll}
        \epsilon \ge \sum_{i=1}^h{\epsilon \times \omega_i} \text{ , where } \epsilon_i = \epsilon \times \omega_i \\ \\
        %\epsilon = \epsilon \times \sum_{i=1}^h{\omega_i} \\ \\
        1 \ge \sum_{i=1}^h{ \omega_i}
    \end{array}
\end{equation}

According to Equation \ref{eq:ec2}, we need to define a weight for each level and ensure that the sum of weights does not exceed $1$. To define such a distribution, we can opt for the following convergent series \cite{zeidler2004oxford}:
\begin{equation}\label{eq:ec21}
\begin{array}{ll}
\omega_i = \frac{1}{i(i+1)} \text{ with positive integers } i \\ \\

\sum_{i=1}^{\infty} \omega_i = 1
\end{array}
\end{equation}

The issue in this case is that most of the allocation will be given to the very first few levels of the decomposition (i.e. bigger weights at the start) because such series decays quickly. Thus, it may lead to injecting too much noise early in the decomposition and cause the \textit{Early Convergence} discussed in Section \ref{sec:ps} due to to the budget being consumed quickly (big $\omega_i$ early on). 

Simply considering a slower series may  not be sufficient. So to slow down the weights distribution, we considered  the sum of multiple instances of the series term by term. For instance, given $S = 1/i(i+1)$, then we define  $S^2 = 1/i(i+1) + 1/i(i+1) = 2/i(i+1)$. Here, $S^2$ decays slower than $S$, but it no longer satisfies the condition in Equation \ref{eq:ec2} and the sum of weights is no longer equal to 1. However, we notice that the first term of $S^2$ is $S^2_1 = 2/(1+1) = 1$, so the remaining terms have the sum equal to 1: $\sum_{i=2}^n{2/i(i+1)} = 1$. Therefore, we can offset the start of the series (in the case of $S^2$ the \textit{offset} is $1$, we refer to it as $os$) to obtain a new series that: (1) decays slower and (2) satisfies Equation \ref{eq:ec2}. 

So in the case of $S = 1/i(i+1)$, we can use a series $S^k = k/i(i+1)$ with a given \textit{offset} $os$ to define the weight $\omega_i$ as:

\begin{equation}\label{eq:w_i}
    \omega_i  = S^k_{(i + os )}
\end{equation}

Thus, we define a function $get\_weight(.)$ that will be used by $Secure\_cc$ and $Secure\_ss$ functions (used in Algorithm \ref{alg:over}) to obtain the proper budget for an operation based on the block's depth at each phase of Algorithm \ref{alg:over}.

% Setting the adequate \textit{offset} $os$ depends on the $S$, in this work we considered  $S = 1/n(n+1)$ but any other series can be used. To find $os$ given this definition of $S$, we need to find $os$ for which  the following condition holds : $\sum_{i=os}^n{k/i(i+1)} \leq 1$. Given $S$, we know that $\sum_{i=1}^n{1/i(i+1)} \leq 1-(1/(n+1))$ which implies $\sum_{i=1}^n{k/i(i+1)} \leq k(1-(1/(n+1)))$. Thus, $os$ must satisfy:  $\sum_{i=1}^{os}{k/i(i+1)} \ge k-1 $ which implies : $k(1-(1/(os+1))) \ge k-1$. So, for any given $S^k$, the \textit{offset} $os$ must satisfy the following: $os \ge k-1$ in order to guarantees the sum of the weights is $\leq 1$ .
Setting the appropriate \textit{offset} \(os\) depends on the choice of \(S\). In this work, we consider \(S = \frac{1}{i(i+1)}\), though other series can also be used. To determine \(os\) for this definition of \(S\), we need to ensure that the following condition holds:

Given \(S\), we know that:
\[
\sum_{i=1}^n \frac{1}{i(i+1)} \leq 1 - \frac{1}{n+1},
\]
which implies:
\[
\sum_{i=1}^n \frac{k}{i(i+1)} \leq k \left( 1 - \frac{1}{n+1} \right) \le k
\]
Thus, we define \(os\) such as:
\[
\sum_{i=1}^{os-1}\frac{k}{i(i+1)} + \sum_{i=os}^{n} \frac{k}{i(i+1)} \leq k
\]
And in such a way :
\[
        \sum_{i=1}^{os-1}\frac{k}{i(i+1)}  \ge k-1 \text{ and }  \sum_{i=os}^{n} \frac{k}{i(i+1)} \le 1
\]
The first sum $\sum_{i=1}^{os-1}\frac{k}{i(i+1)}$ represents the sum of terms that we will ignore, and the second sum $\sum_{i=os}^{n} \frac{k}{i(i+1)}$ represents the sum of terms that we want to use in Equation \ref{eq:w_i}. Based on this, we get:
\[
\sum_{i=1}^{os-1} \frac{k}{i(i+1)} \geq k-1 \implies k \left( 1 - \frac{1}{(os-1)+1} \right) \geq k-1
\]
which leads to the condition, for any given \(S^k\), the \textit{offset} \(os\) must satisfy:
\[
os \geq k
\]
to ensure that the sum of the weights remains within the bound \( \leq 1 \).

\begin{figure}[!h]
    \centering
    \includegraphics[width=1\linewidth]{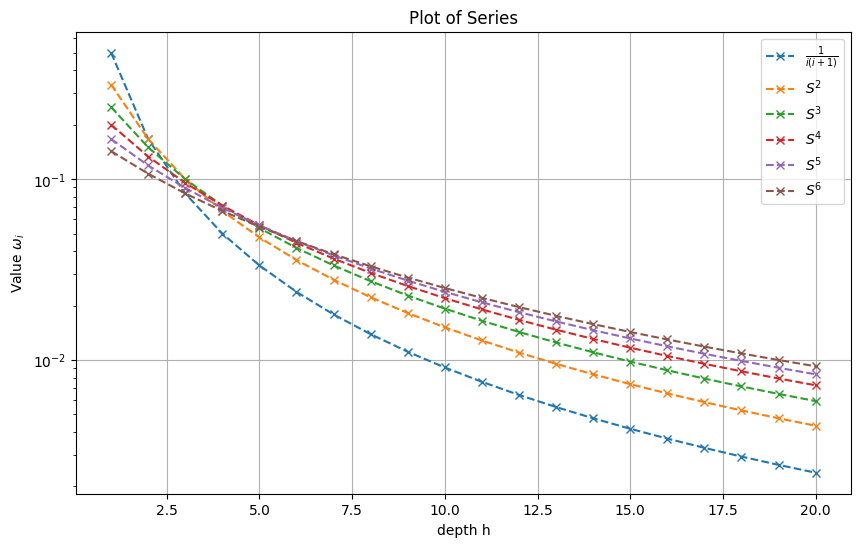}
    \caption{$\omega_i$  for each level of depth $h$ using $S^k$}
    \label{fig:decay_series}
\end{figure}

To show the difference in decay between $S$ and its variants $S^k$, Figure \ref{fig:decay_series} illustrates the $\omega_i$ values obtained at each depth using different series.

Other alternatives, such as uniform budget distribution, are not possible since $h$ is not known in advance. Additionally, the use of a growing series might be more interesting than a decaying series \cite{cormode} in some cases, but it also requires a predefined depth $h$. Without a predefined $h$, it would be very difficult to create such a series while ensuring that the constraint in Equation \ref{eq:ec2} holds.

\subsection{Secure Convergence and Partitioning}\label{sec:cc_ss}
In Algorithm \ref{alg:over}, we showed that both phases use the same functions for convergence and splitting, differing only in the set of parameters, which hints that they follow the same logic. 

\subsubsection{Secure convergence condition $Secure\_cc$:} 
Algorithm \ref{alg:s_cc} describes the function $Secure\_cc$, which takes as input a block $B$, a metric based on which to determine convergence, a threshold $\theta$, and $\epsilon$, the privacy budget allocated to the convergence in the current phase ($\epsilon^1_{cc}$ or $\epsilon^2_{cc}$ in Algorithm \ref{alg:over}).

Algorithm \ref{alg:s_cc} starts with computing the budget $\epsilon_i$ corresponding to the current iteration based on Equations \ref{eq:ec2} and \ref{eq:w_i} and the depth of the block (Algorithm \ref{alg:s_cc} Lines 1-2).  The function $depth(.)$ returns the depth of the block in parameter based on the current decomposition phase. When a block $B$ moves from phase 1 to phase 2 of Algorithm \ref{alg:over}, this function resets the depth of $B$ to 1.

Then, $Secure\_cc$ computes and perturbs the desired metric (Algorithm \ref{alg:s_cc} Line 3). Finally, it returns True or False on whether the perturbed metric is less than or equal to $\theta$. For this function the choice of $\theta$ does not have much affect on the output of the algorithm (as long as $\theta$ is small) as mentioned by \cite{privtree}, so in our algorithm this parameter is set to $0$ by default.

\begin{algorithm}[H]
\caption{Secure\_cc}\label{alg:s_cc}
\begin{algorithmic}[1]
\Require $B$: block, metric,$\epsilon,\theta$

\State  $\omega_i  \gets get\_weigth(depth(B))$
\State $\epsilon_i \gets \omega_i \times \epsilon$
\State $value \gets metric(B) + Lap(\Delta_{metric}/\epsilon_i)$
\State $\textbf{return } value \leq \theta$

\end{algorithmic}
\end{algorithm}

The $Secure\_cc$ uses the \textit{Laplace mechanism} (Algorithm \ref{alg:s_cc} Line 3) to ensure DP, and the scale of the noise depends on $\Delta_{metric}$, the sensitivity of the $metric$. The sensitivity of the metric will be discussed in details in Section \ref{sec:sensi}.

\subsubsection{Secure splitting strategy $Secure\_ss$:} 
The Algorithm \ref{alg:s_ss} illustrates all the steps of the function $Secure\_ss$. This function takes as parameters a block, metric, and the budget $\epsilon$ ($\epsilon^1_{ss}$ or $\epsilon^2_{ss}$ ) allocated for $SS$ in the current phase of Algorithm \ref{alg:over}. $Secure\_ss$ starts by computing the budget for the current iteration (Algorithm \ref{alg:s_ss} Lines 1-2), in a similar manner as the $Secure\_cc$. Then, it loops over all the possible cutting points of the block and calculates a sampling probability (Algorithm \ref{alg:s_ss} Lines 3-5) based on the Definition \ref{def:em} of the \textit{Exponential mechanism}.

The function $score$ takes as a parameter $B$, $metric$, and a cutting point $i$. The function $score$ will treat $B$ as if it were split at point $i$ into two halves, $B_L$ and $B_R$. According to the $metric$ (in other words, according to the phase of Algorithm \ref{alg:over}), it will give the following score to the point $i$:

\begin{itemize}
    \item \textbf{Case metric is '$Min$'}: $Score_i = - \min\{\min_{B_{L}}, \min_{B_{R}}\}$. Where $\min_{B_{L}}$ (or $\min_{B_{R}}$) is the minimum between the number (count) of empty and nonempty cells in the left block (respectively in the right block). Thus, the cutting point with $\min \approx 0$ will get the highest score as it leads to a perfect split of cells.
    \item \textbf{Case metric is '$AE$'}: $score_i = - (AE_{B_{L}} + AE_{B_{R}})$. Thus, giving the highest score to the cutting point that provides a split with minimal $AE$ .
\end{itemize}

\begin{algorithm}[H]
\caption{Secure\_ss}\label{alg:s_ss}
\begin{algorithmic}[1]
\Require $B$: block, metric,$\epsilon$

\State  $\omega_i  \gets get\_wiegth(depth(B))$
\State $\epsilon_i \gets \omega_i \times \epsilon$
\For {$ i \in  \Omega_B$}
		\State $prob[i] \gets  exp\left( \frac{\epsilon_i \times score(B,metric,i)}{4 \times \Delta_{metric}} \right) $
\EndFor
\State $point \gets random\_choice(\Omega_B,prob)$
\State $B_L,B_R \gets split(B,point)$
\State $\textbf{return } B_L,B_R$

\end{algorithmic}
\end{algorithm}

Since we are computing the measure twice (on $B_L$ and $B_R$), notice that we set the sensitivity of the score as $ \Delta_{score} = 2 \times \Delta_{metric}$ (The sensitivity of the metric will be discussed in details in Section \ref{sec:sensi}).

Afterwards, it applies \textbf{unequal probability sampling} (line 6) to select a random cutting point and splits $B$ into $B_L$ and $B_R$, which are returned as results.\\

\subsubsection{Sensitivity Min, Sum and AE}\label{sec:sensi}

In order to apply the \textit{Laplace mechanism} and \textit{Exponential mechanism} used in $Secure\_cc$ and $Secure\_ss$ respectively, we need to compute the sensitivity of each metric used.
\begin{theorem}[Sensitivity of the Min, Sum and AE]\label{th:sen_s_ae}
For any neighbouring tensors $T$ and $T'$: $\Delta_{Min} = 1$, $\Delta_{Sum} = 1$ and $\Delta_{AE} = 2$.
\end{theorem}

Proof:

\textbf{In the case of the $Sum$ (and the $Min$)}: It is straightforward, $\Delta_{Sum} = 1$ ($\Delta_{Min} = 1$), since given two neighboring tensors $T$ and $T'$, the $Sum$ ($count$ of empty or nonempty cells) will change by $1$ because of a cell having $+1$ or $-1$.

 \textbf{In the case of $AE$}: Consider $T$ and $T'$. In $T'$, suppose the cell $c_l$ got $+1$, we compute the $AE$ as follows:
 \begin{equation}\label{eq:se_ae}
    \begin{array}{ll}
        AE(T) = \sum_{c \in \Omega_D} {\vert \Bar{x} - c \vert} \text{ and,} &  \Bar{x} = \frac{Sum(T)}{\vert \Omega_D \vert} \\
         AE(T') = \sum_{c \in \Omega_D, c\neq c_l } {\vert \Bar{x'} - c \vert + \vert \Bar{x'}- c_l \vert}  \text{ and,} & \Bar{x'} = \frac{Sum(T) + 1}{\vert \Omega_D \vert}
    \end{array}
\end{equation}
Based on Equation \ref{eq:se_ae}, $AE(T') \leq \sum_{c \in \Omega_D} {\vert \Bar{x} - c \vert} + 2 \times \frac{n-1}{n}$ which implies that $\Delta_{AE} = \vert AE(T) - AE(T')\vert \leq 2 \times \frac{n-1}{n} \leq 2$.

\subsection{Privacy budget}
To compute the privacy budget consumption of our decomposition algorithm, and since each branch manipulates a distinct portion of tensor $T$ (as mention in Section \ref{sec:ps}), we need only to track the consumption of only one path from the root $T$ to a leaf node. Then based on the parallel composition, we can deduce the total consumption. 

Each leaf node published by our algorithm must be produced by a branch that extends from the root and goes through two phases of decompositions as described in Algorithm \ref{alg:over}. Based on this, let's assume that a branch went $k$ iterations in our first decomposition phase and $p$ iterations in the second phase. In the last step, the algorithm perturbs the leaves before publishing them.  Given that all these iterations are applied consecutively to the same portion of the tensor, the total privacy budget is computed using the sequential composition property of DP, accumulating the budget consumed in each iteration.

So, to ensure that the budget consumption by the algorithm is in bound of the total budget $\epsilon$ given as input to Algorithm \ref{alg:over}, means proving that for any given leaf block $B_{l}$ this equation holds:
\begin{equation}
    \frac{Pr[B_l \vert T]}{Pr[B_l \vert T']} \leq e^{\epsilon}
\end{equation}

We note $B_l^i$ the $i$'th predecessor of $B_l$ starting from the root and belonging to the path from root to $B_l$. According to our Algorithm \ref{alg:over},  for the $Secure\_cc$ during the first the phase we have the following budget consumption :
\[
    \left| \prod_{i=1}^{k-1}{ \frac{Pr[Sum(B^i_l) + Lap(\lambda_i) > \theta_!]}{Pr[Sum(B'^i_l) + Lap(\lambda_i) > \theta_1]}} \right| \times \frac{Pr[Sum(B^k_l) + Lap(\lambda_i) \leq \theta_1]}{Pr[Sum(B'^k_l) + Lap(\lambda_i) \leq \theta_1]}
\]
And for the $Secure\_ss$ during the first phase:
\[
          \prod_{i=1}^{k}{ \frac{Pr[point_j \vert B^i_l, Min]}{Pr[point_j \vert B'^i_l, Min]}} 
\]
Where $Pr[point_j|B^i_l, metric]$ is the probability $point_j$ being selected as a cutting point given a block $B_l^i$ and a scoring function using $metric$. For the second phase we have a similar budget consumption. For the $Secure\_cc$ in the second phase:
\[
\left| \prod_{i=k}^{p-1}{ \frac{Pr[AE(B^i_l) + Lap(\lambda_i) > \theta_2]}{Pr[AE(B'^i_l) + Lap(\lambda_i) > \theta_2]}} \right| 
         \times \frac{Pr[AE(B^p_l) + Lap(\lambda_i) \leq \theta_2]}{Pr[AE(B'^p_l) + Lap(\lambda_i) \leq \theta_2]}
\]
And for the $Secure\_ss$:
\[
   \prod_{i=k}^{p}{ \frac{Pr[point_i \vert B^i_l, AE]}{Pr[point_i \vert B'^i_l, AE]}} 
\]
Finally, Algorithm \ref{alg:over} perturbs the mean of the leaf:
\[
     \frac{Pr[mean(B_l) + Lap(\lambda) = S ]}{Pr[mean(B'_l) + Lap(\lambda) = S ]} 
\]
Since each operation ($Secure\_cc$ and $Secure\_ss$) is allocated a  dedicated budget for each phase of Algorithm \ref{alg:over} lines 1-4, and according to the budget distribution we proposed in Section \ref{sec:bd} and the definitions in\ref{def:lap},\ref{def:em}  : $Secure\_cc$ will not exceed $\epsilon^1_{cc}$ and $\epsilon^2_{cc}$ in both phases. Similarly, $Secure\_ss$ will not exceed $\epsilon^1_{ss}$ and $\epsilon^2_{ss}$ in both phases. So based on the sequential composition, we have:
\[
     \frac{Pr[B_l \vert T]}{Pr[B_l \vert T']} \leq e^{\epsilon^1_{cc}} \times e^{\epsilon^1_{ss}} \times e^{\epsilon^2_{cc}} \times e^{\epsilon^2_{ss}} \times e^{\epsilon_p}  \leq e^{\epsilon}
\]
% \begin{equation} \ref{eq:pa_f_f}
%      \begin{array}{ll}
%           \frac{Pr[B_l \vert T]}{Pr[B_l \vert T']} \leq \exp^{\epsilon_{cc}^1} \times \exp^{\epsilon_{ss}^1} \\
%           \times \exp^{\epsilon_{cc}^2} \times \exp^{\epsilon_{ss}^2} 
%           \exp^{\epsilon_{p}}\\
%           \frac{Pr[B_l \vert T]}{Pr[B_l \vert T']} \leq exp^{\epsilon_{cc}^1 + \epsilon_{ss}^1 +\epsilon_{cc}^2 +\epsilon_{ss}^2 + \epsilon_{p}}\\

%           \implies \frac{Pr[B_l \vert T]}{Pr[B_l \vert T']} \leq \exp^{\epsilon}
%      \end{array}
% \end{equation}
This result shows that our algorithm doesn't exceed the privacy budget given as an input, independently of the depth $h$ of any branch during the decomposition. Based on the \textit{Parallel Composition} property of DP, this conclusion can be generalized to all paths of the decomposition.

\section{Experiments}\label{sec:eval}
In order to assess the performance of our proposed algorithm \texttt{RIPOST}, we conducted an extensive comparative experiment with existing algorithms on several different datasets. In this section, we discuss the main results and observations we obtained.
\subsection{Experiments Setup}
{\noindent\bf Datasets.} In our experiments, we used the following well-known datasets:  

(i) \textbf{Census Adult}\footnote{\url{http://archive.ics.uci.edu/ml/datasets/Adult}} is a benchmark dataset containing demographic information about individuals and their income.  

(ii) \textbf{Fire Department and Emergency Medical Services Dispatched Calls for Service (Fire)}\footnote{\url{https://data.sfgov.org/Public-Safety/Fire-Department-and-Emergency-Medical-Services-Dis/nuek-vuh3/data}} contains records of emergency calls made to the fire department in San Francisco.  

(iii) \textbf{Gowalla-2D}\footnote{\url{http://snap.stanford.edu/data/loc-Gowalla.html}} is a geo-location check-in dataset.  

(iv) \textbf{Jm1}\footnote{\url{https://www.openml.org/search?type=data&sort=runs&id=1053&status=active}} is a dataset containing static source code analysis data for defect detection.  

For each dataset we generated a several tensors and thousands of queries, the Table \ref{tab:stats} shows the different stats on each dataset. In the remainder of this section, we will emphasize how we used each dataset/workload.
\begin{table}[h!]
    \centering
    \small
    \begin{tabular}{|c|c|c|c|c|} \hline 
         \textbf{Datasets} & \textbf{\#Dimensions} &  \textbf{\#Workloads}&  \textbf{\#Queries}& \textbf{Domain}\\ \hline 
          Adult&15&  119 + 8&  $(5  + 8) \times 3000$& $8.9 \times 10^{18}$\\ \hline 
          Fire&35&  119 + 8&  $(5+8) \times 3000$& $2.291 \times 10^{57}$\\ \hline 
          Gowalla&2&  1&  $3000$& $10^6$\\ \hline 
          Jm1&21&  456&  $5 \times 3000$& $2 \times 10^{21}$\\\hline
    \end{tabular}
    \caption{Dataset and workload statistics}
    \label{tab:stats}
\end{table}
% \vspace{-1}
{\noindent\bf Evaluation metrics.} 
In our evaluation, we used the RMSE (Root-Mean-Square Error) to measure the error between the query answer from the original dataset and those from the perturbed view. 
The RMSE is defined as: $\text{RMSE} = \sqrt{\frac{1}{n} \sum_{Q\in W} (Q(T) - Q(T'))^2}$

where $W$ is the workload, and $T$, $T'$ are the original and private views of the tensor, respectively (regardless of the method used to create $T'$).  
To compare the results across different algorithms, we calculated the relative RMSE, denoted as \textbf{\textit{R-RMSE}}, by dividing the RMSE obtained by each approach on a workload $W$ by the RMSE obtained using our approach on the same dataset.  

Thus, if $\text{R-RMSE} > 1$, it means our approach outperformed the other, and \textbf{\textit{R-RMSE}} indicates the magnitude of the difference. Conversely, if $\text{R-RMSE} < 1$, it means our approach underperformed compared to the other, and \textbf{\textit{R-RMSE}} indicates the fraction that the RMSE of the other approach represents relative to ours.

\medskip
{\noindent\bf Competitors.}  
We compared \texttt{RIPOST} with several recent solutions (in addition to the naive solution \textit{Identity}\cite{iden}) from the literature (see Section \ref{sec:rw}), namely:  
(i) HDPView \cite{hdpview} and PrivTree \cite{privtree} for partitioning approaches;  
(ii) HDMM \cite{hdmm} and DAWA \cite{dawa} for workload-dependent approaches;  
(iii) PrivBayes \cite{privbayes} and P3GM \cite{p3gm} for generative models creating DP datasets.  
\medskip  

{\noindent\bf Hyperparameters.}  
In all experiments, we set the total privacy budget to $\epsilon = 0.1$. For \texttt{RIPOST}, we set $\alpha = 0.3$, $\gamma = 0.9$, and $\beta = 0.4$. A complete analysis of the effect of each of these hyperparameters on the performance of \texttt{RIPOST} is provided in Section \ref{sec:sen_ana}. For the other solutions, we used the same hyperparameter values as those reported in their original experiments.  
A disclaimer: The model presented in \cite{p3gm} computes the privacy budget based on both the data size and the training sample. As a result, it was quite complicated to tune the $\epsilon$ of this model for all the tensors generated in our tests, with minimal change on setup of the original paper. So to ensure fairness to all competitors, \cite{p3gm} was allocated a larger budget within the range of [0.1, 1], depending on the tensor.\medskip 

{\noindent\bf Source Code:} The code for all the experiments conducted in this paper is available on GitHub\footnote{\label{note1}\url{https://github.com/AlaEddineLaouir/RIPOST.git}}. In addition to the experiments presented here, additional experiments can be found in the GitHub repository (due to space limitations).\medskip 

\subsection{\texttt{RIPOST} performances analysis}
\subsection*{Partitioning and generative approaches}
In this first part of the experiments, we compared the performance of \texttt{RIPOST} with HDPView, PrivTree, PrivBayes, and P3GM. For these tests, we used the Adult and Fire datasets, and for each, we created a set of 2, 3, ..., 6-dimensional tensors, each with its own workload. Table \ref{tab:stat2} shows the number of distinct tensors/workloads generated and used for each number of dimensions.  

\begin{table}
    \centering
    \begin{tabular}{|c|c|c|c|c|c|} \hline 
         \# Dimensions & 2 & 3 & 4 & 5 & 6 \\ \hline 
         \# Tensors/Workloads & 21 & 35 & 35 & 21 & 7 \\ \hline
    \end{tabular}
    \caption{Number of Tensors/Workloads per Number of Dimensions}
    \label{tab:stat2}
\end{table}

These different tensors/workloads are used to ensure a thorough and robust evaluation.
\begin{figure}
    \centering
    \includegraphics[width=1\linewidth]{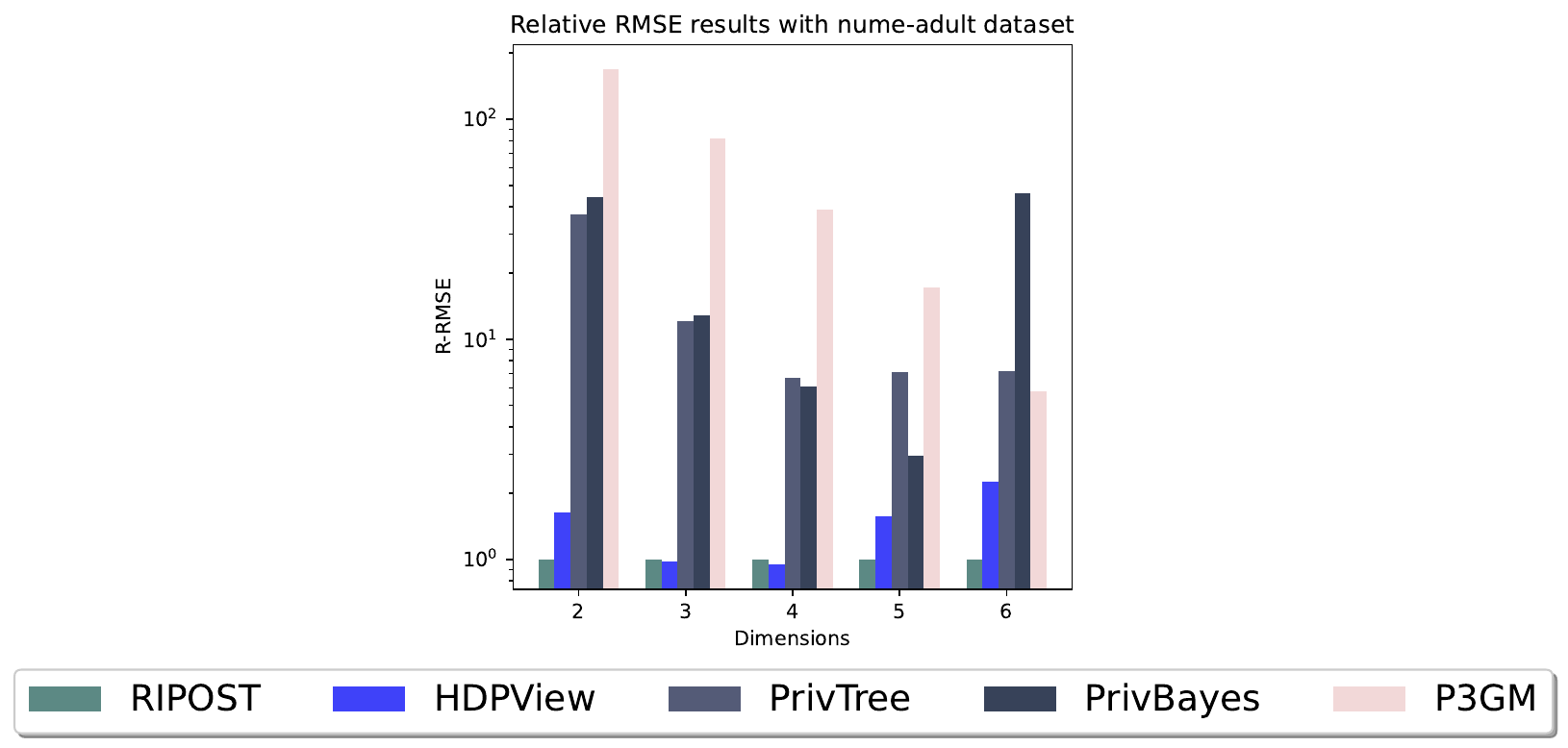}
    \caption{R-RMSE results based on Adult dataset}
    \label{fig:rmse-adult}
\end{figure}
\begin{figure}
    \centering
    \includegraphics[width=1\linewidth]{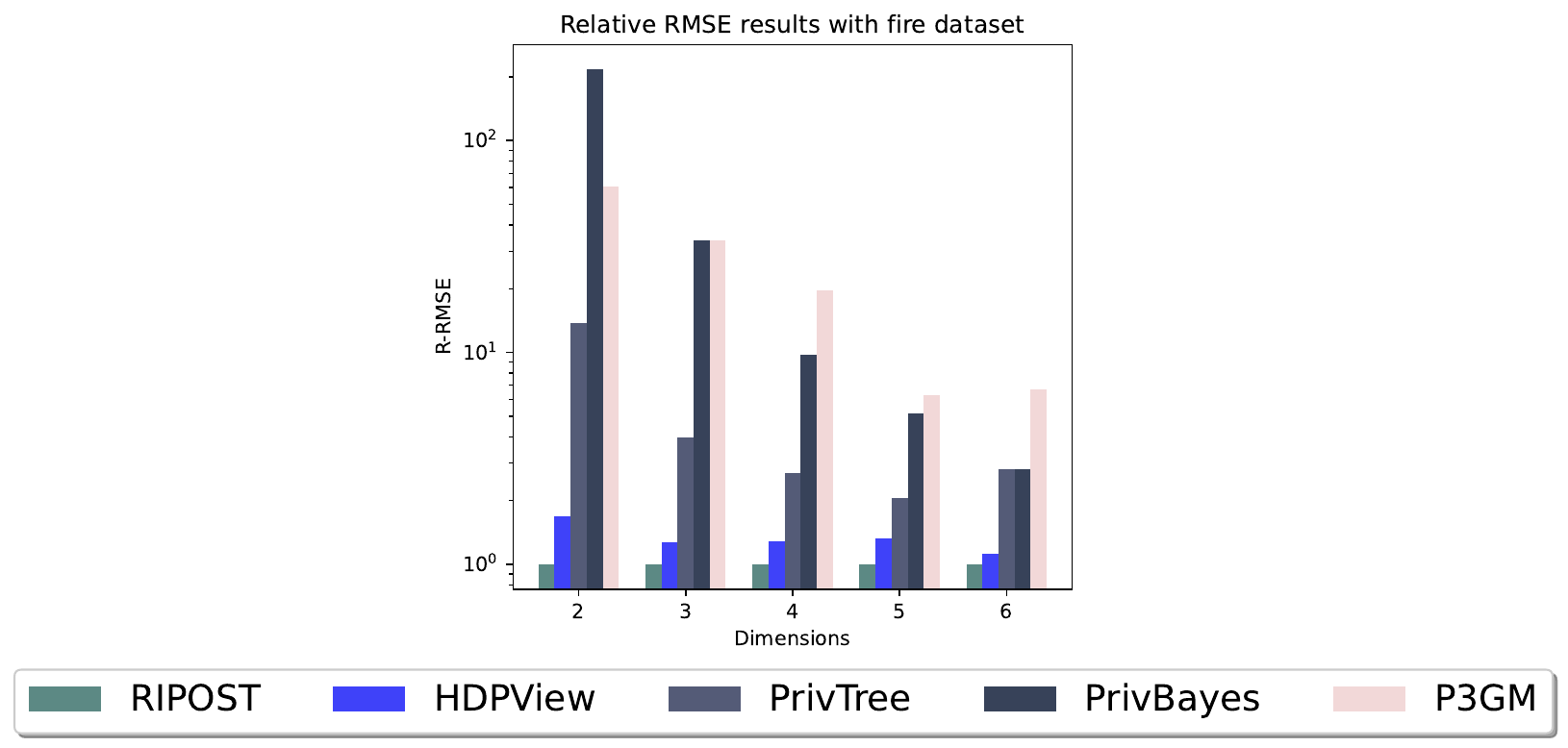}
    \caption{R-RMSE results based on Fire dataset}
    \label{fig:rmse-fire}
\end{figure}
In Figures \ref{fig:rmse-adult} and \ref{fig:rmse-fire}, we show the average R-RMSE obtained by each approach on both the Adult and Fire datasets, respectively. First, we notice that our approach is able to maintain better performance than all the others in the different tests. In Figure \ref{fig:rmse-adult}, we see that our closest competitor is HDPView, but its performance pales as the number of dimensions grows. This is due to the large number of blocks generated at high-level tensors, but HDPView fails to calibrate the PE and AE compared to \texttt{RIPOST}. Also, in both Figures \ref{fig:rmse-adult} and \ref{fig:rmse-fire}, we notice that PrivTree has worse performance than \texttt{RIPOST} and HDPView due to its limitations, as mentioned in Section \ref{sec:ps}, such as data-independent SS.  

Regarding the generative approaches, PrivBayes and P3GM, we notice that their performance improves with higher dimensions due to the increase in training data. However, this does not make them outperform the partitioning approaches at any point in our tests, and this is due to the fact that they are not optimized for tensors and OLAP range queries as tasks.  In particular, these models does not  take into account the \textit{measure} attribute created due to the aggregation (as explained in Section \ref{sec:pre}).

To further test the performance of these approaches, we scaled the tests to higher dimensions. We used the Adult dataset to create tensors with dimensions ranging from 7 to 14, each with a workload of 3000 queries to test the quality of the generated views.  

\begin{figure}
    \centering
    \includegraphics[width=1\linewidth]{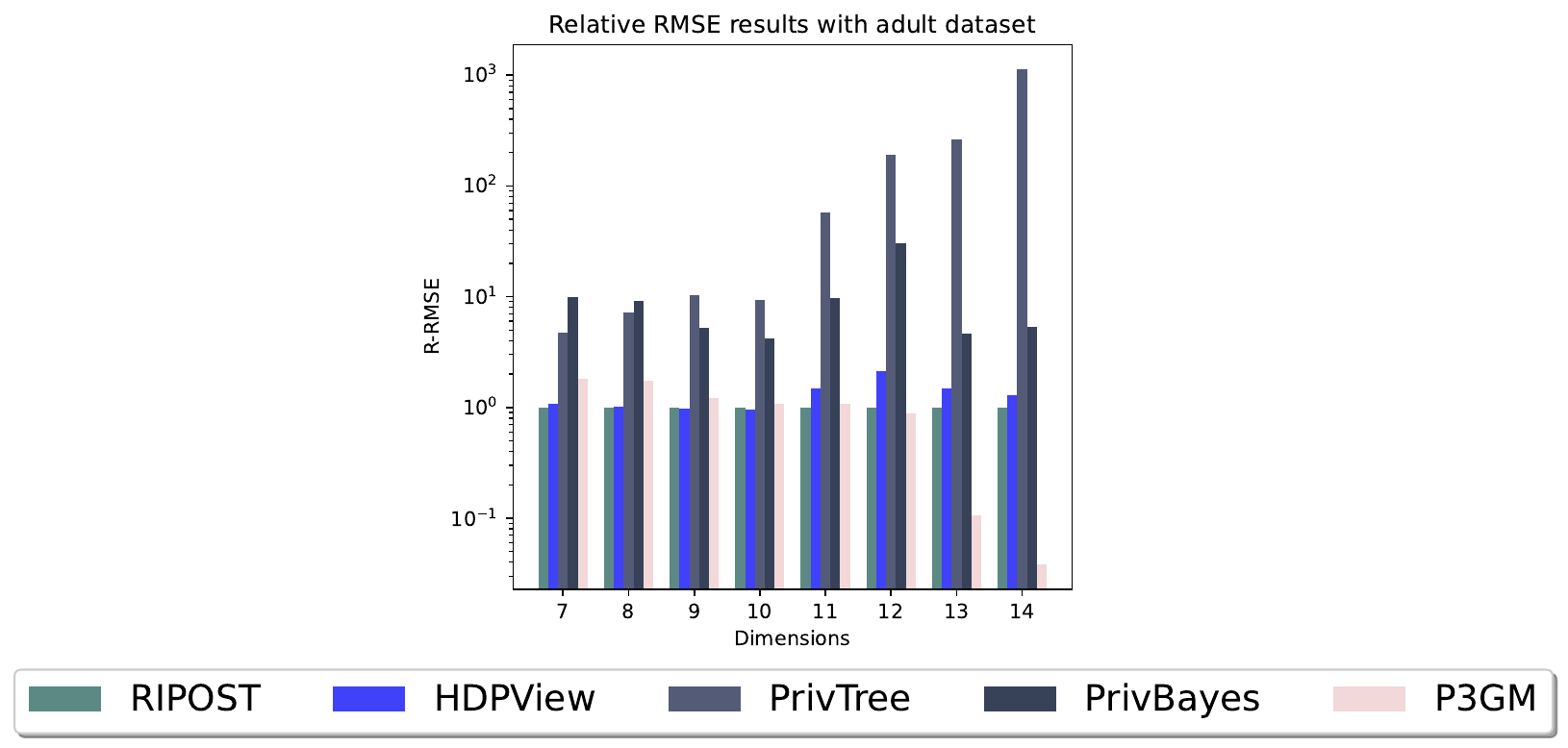}
    \caption{R-RMSE on higher dimension tensors based on the Adult dataset.}
    \label{fig:rmse_scala}
\end{figure}
Based on the results shown in Figure \ref{fig:rmse_scala}, we observe that our approach remains consistent and scales better than the other approaches. Also, HDPView remains a  close competitor, but it is important to note that the generative approaches are also improving their results. Particularly, \texttt{P3GM} had two main advantages: (i) In the case of the \textit{Adult} dataset, creating a tensor with 14 dimensions (out of the 15 available) does not introduce much aggregation, resulting in most values in the \textit{measure} attribute being 1. This significantly reduces the effect of \texttt{P3GM} neglecting this attribute. (ii) The difficulty encountered when tuning its budget led to \texttt{P3GM} consuming more budget than the other methods in these tests, which consequently introduced less noise.

% Particularly P3GM, and this is due two main reasons: (i) in the case of \textit{Adult} dataset, creating a tensor of 14 dimensions (out of 15 available) do not introduce much aggregation thus most of the values in \textit{measure} attribute are 1. Which significatly reduces the effect of P3GM neglacting this attribute. Secondly (ii) the dificulty faced when tununing its budget allowd to consume more than the others in these tests, consequently introducing less noise.

\subsection*{Spatial Data}
Another important use case is the publishing of spatial data, where partitioning algorithms are the most commonly used for this type of data. To highlight the performance of \texttt{RIPOST}, we compared its performance on the Gowalla dataset with HDPView and PrivTree. For this test, we generated 3000 queries to assess the quality of the view created.
\begin{figure}
    \centering
    \begin{minipage}{0.22\textwidth}
        \centering
        \includegraphics[width=\linewidth]{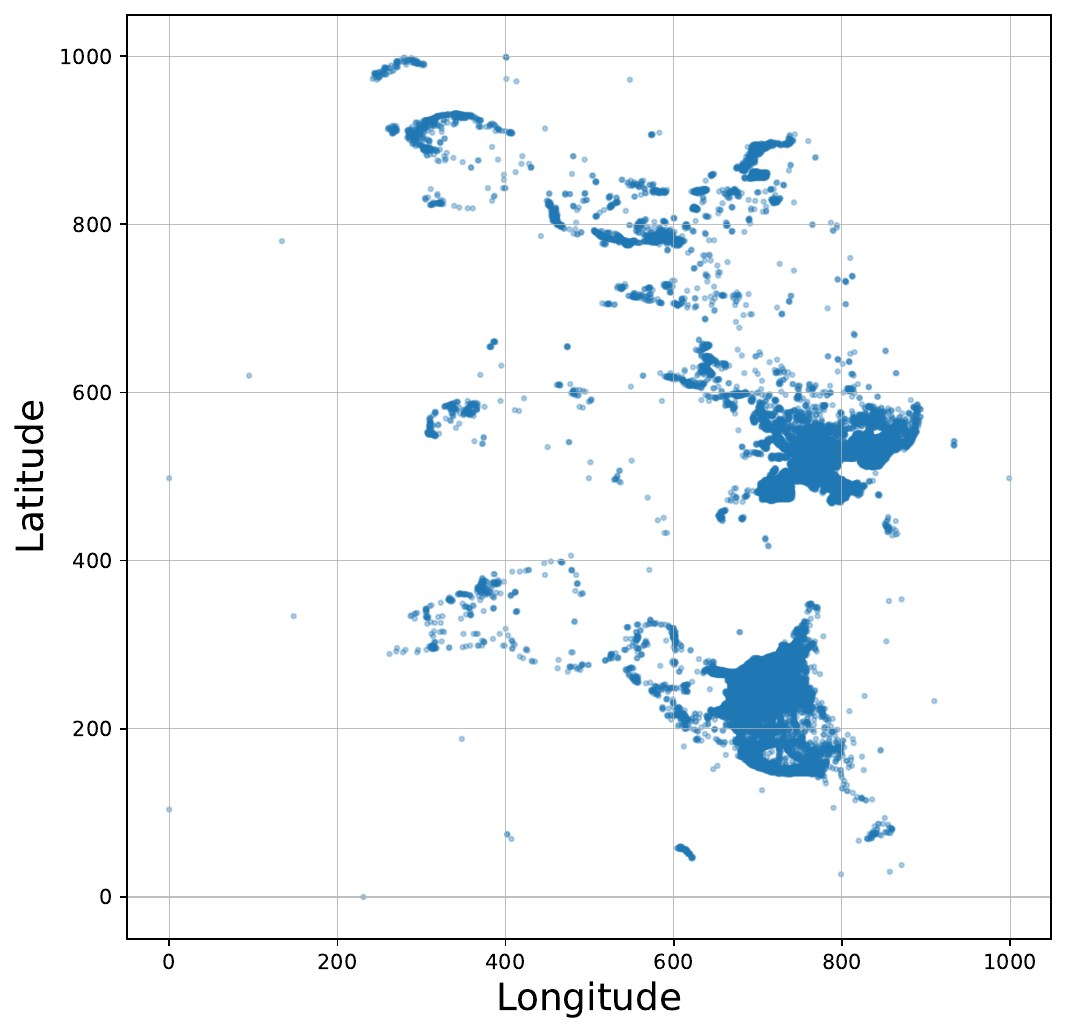}
        \caption{\small Gowalla}
    \end{minipage}%
    % \hspace{0.1\textwidth}
    \begin{minipage}{0.21\textwidth}
        \centering
        \includegraphics[width=\linewidth]{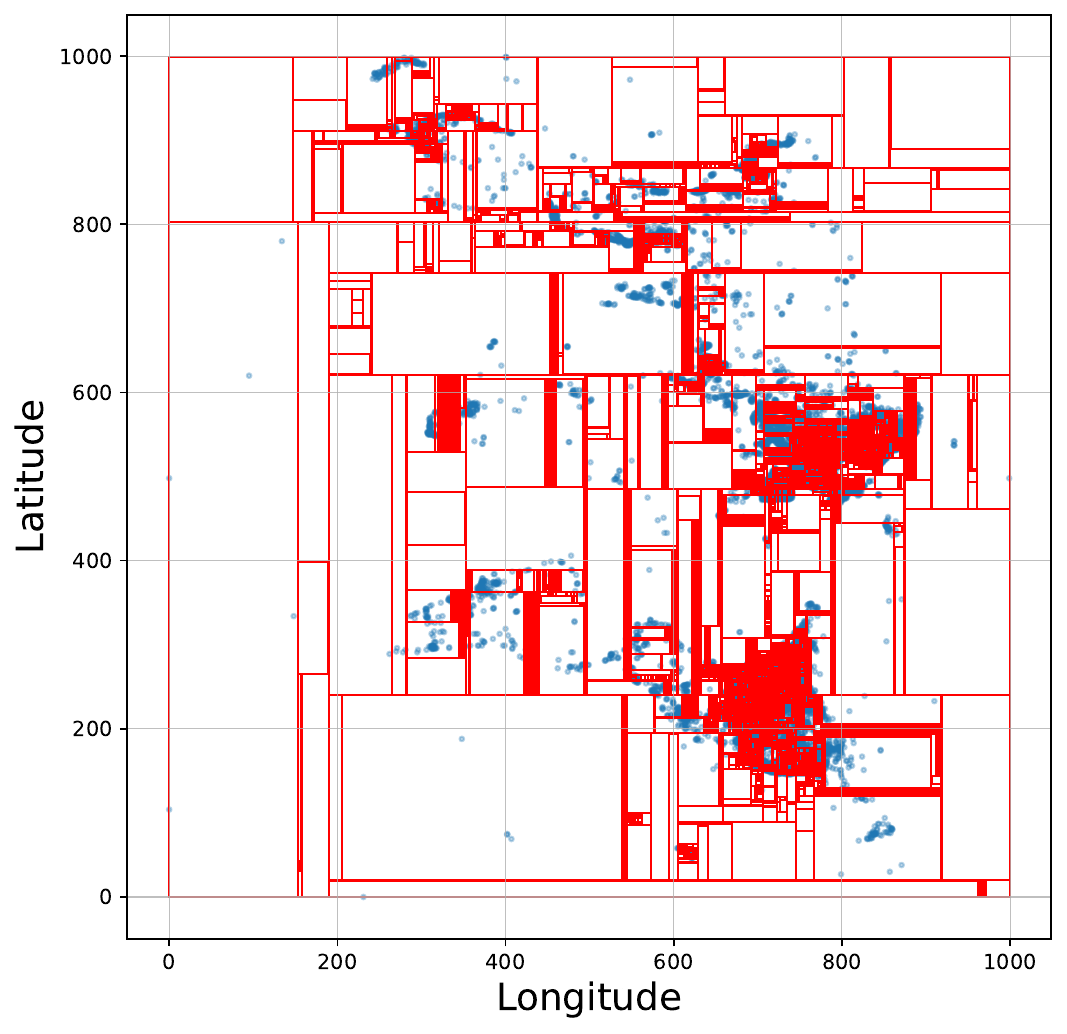}
        \caption{\small \texttt{RIPOST}}
    \end{minipage}
    % \\[1ex] % Line break between rows
    \begin{minipage}{0.21\textwidth}
        \centering
        \includegraphics[width=\linewidth]{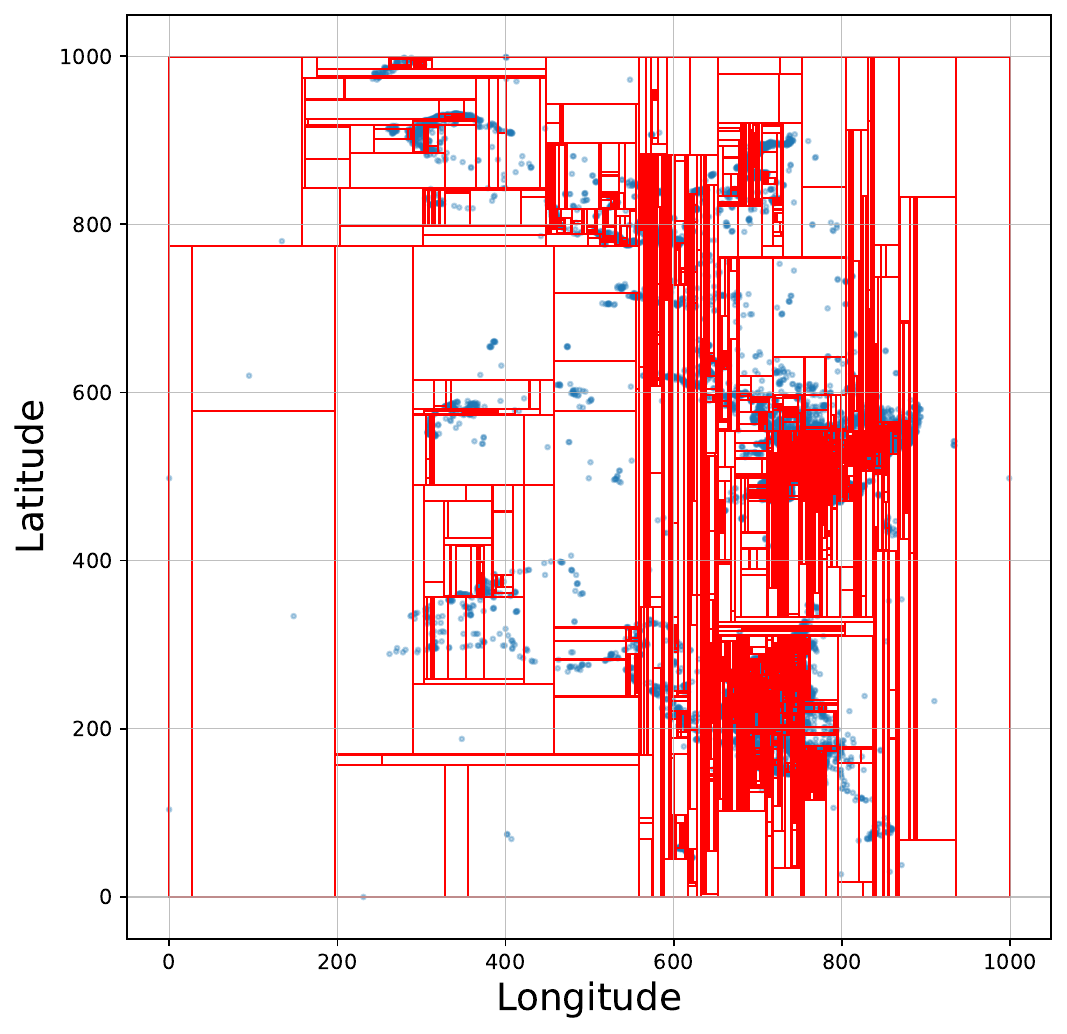}
        \caption{\small HDPView}
    \end{minipage}%
    % \hspace{0.1\textwidth}
    \begin{minipage}{0.21\textwidth}
        \centering
        \includegraphics[width=\linewidth]{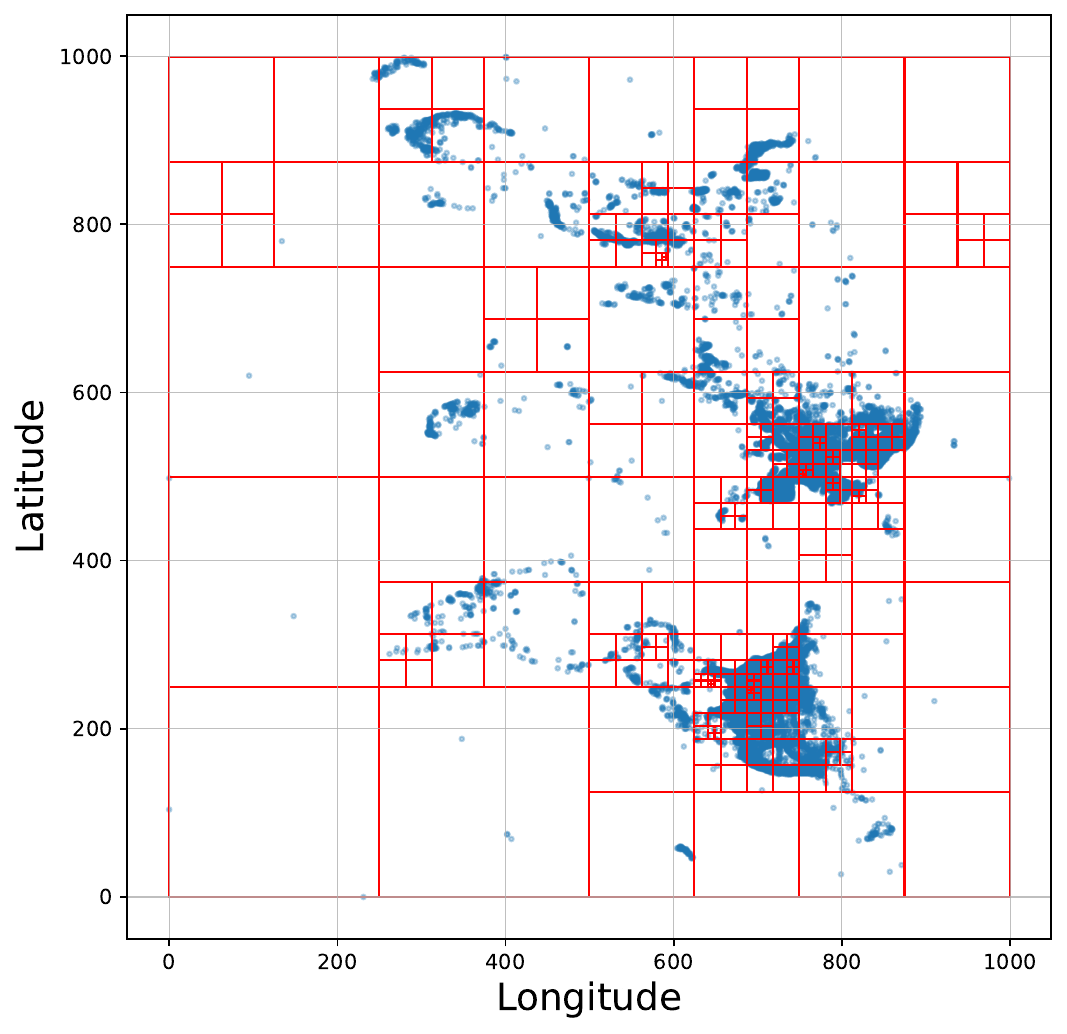}
        \caption{\small PrivTree}
    \end{minipage}
    \caption{Gowalla-2D Partitioning}
    \label{fig:gowalla-par}
\end{figure}
Figure \ref{fig:gowalla-par} shows the spatial partitioning of the Gowalla dataset by \texttt{RIPOST}, HDPView, and PrivTree, respectively. We can easily notice that PrivTree generated the fewest blocks (converging early), further supporting the observations and critiques made in Section \ref{sec:ps}. As a result, it gave an RMSE $\times 25$ larger than ours.

Comparing \texttt{RIPOST} to HDPView, from Figure \ref{fig:gowalla-par}, we can clearly see that \texttt{RIPOST} generated/detected larger empty blocks in different portions of the domain due to the first phase of partitioning. It then focused deeper partitioning on the densely populated portions, thanks to the second phase. Meanwhile, HDPView applied more partitioning across a wider portion of the domain. Partitioning empty regions wrongly and keeping the blocks mixed, did impact the performances of HDPView. In terms of RMSE, and based on our test queries, HDPView gave an RMSE $\times 1.27$ larger than \texttt{RIPOST}.

\subsection*{Workload dependent approaches}
In order to compare \texttt{RIPOST} with HDMM, DAWA, and the naive solution Identity, we used tensors with smaller domains. This is because both HDMM and DAWA have high computational complexity, and Identity requires a large amount of memory storage for big domains, making them impractical for high-dimensional tensors. Therefore, we used the Adult dataset with a pre-selected set of dimensions with a small domain (referred to as \textit{Small-Adult}), and also the Jm dataset. In these experiments, we increased the number of dimensions as much as possible, creating a tensor and 3000 queries for testing at each step.

\begin{figure}
    \centering
    \includegraphics[width=1\linewidth]{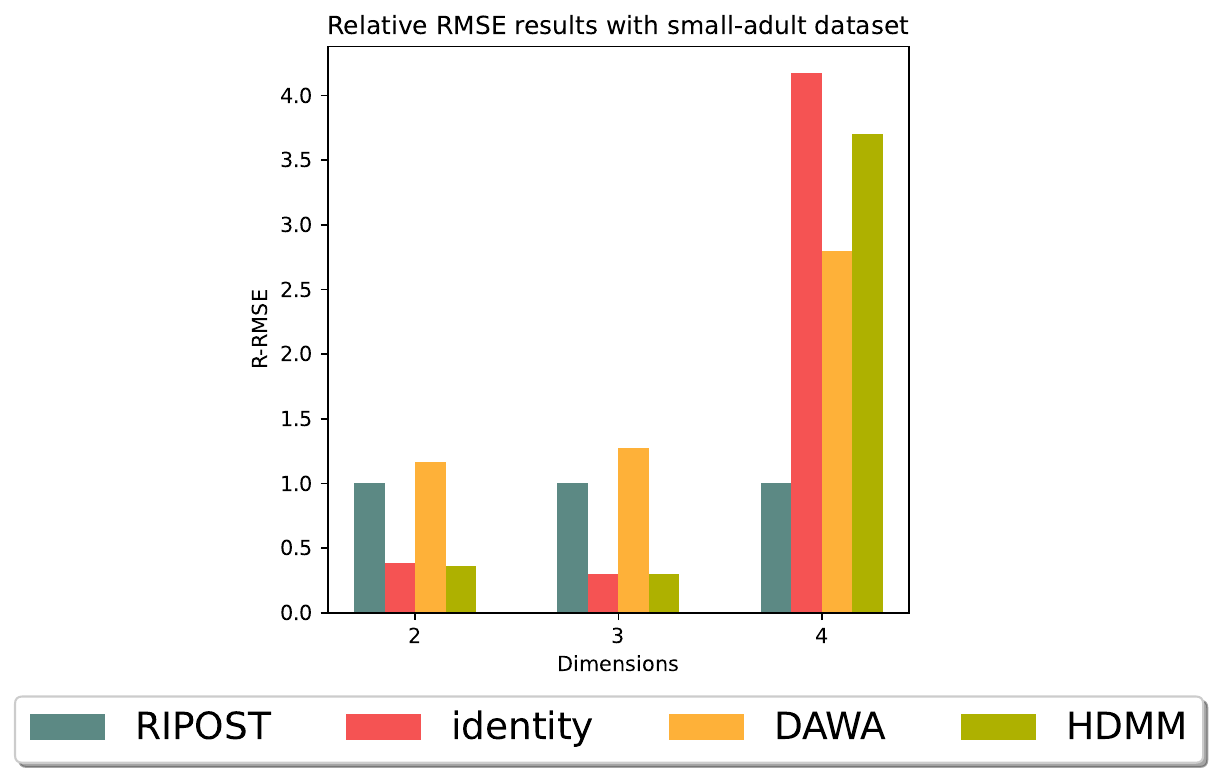}
    \caption{R-RMSE results based on small-adult dataset}
    \label{fig:rmse-smallad}
\end{figure}
\begin{figure}
    \centering
    \includegraphics[width=1\linewidth]{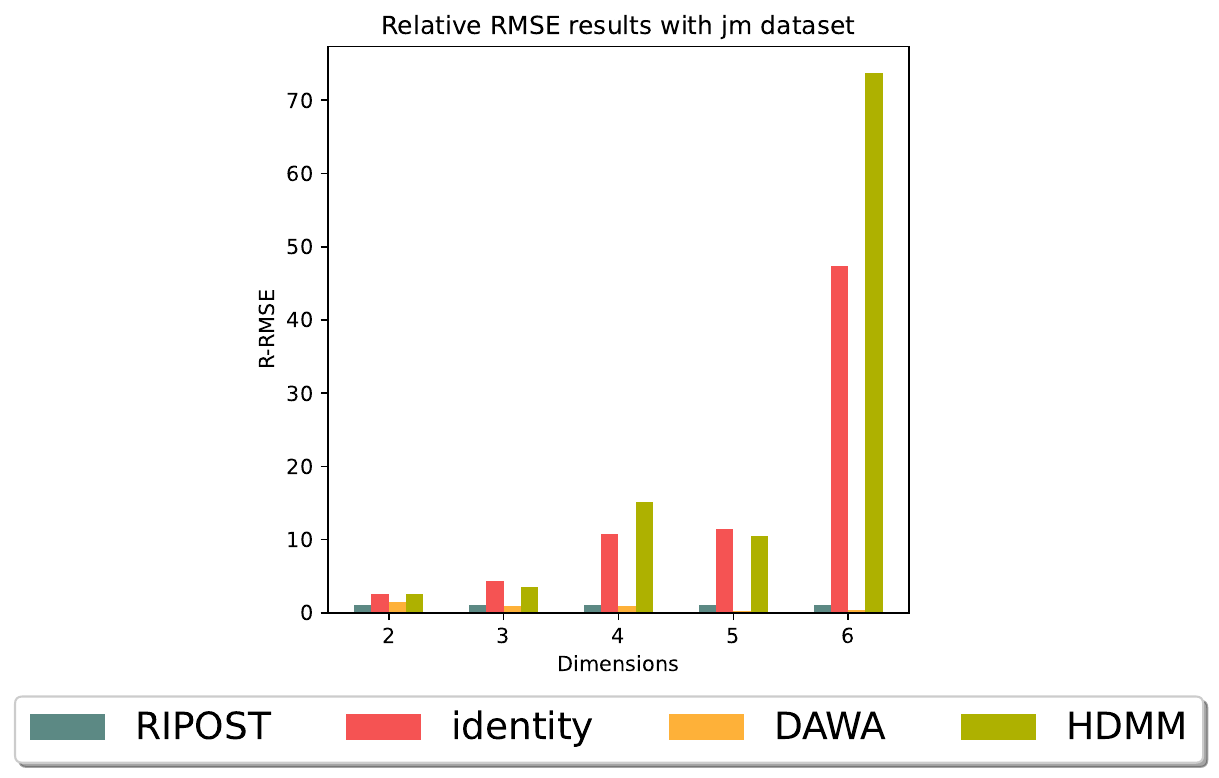}
    \caption{R-RMSE results based on Jm dataset}
    \label{fig:rmse-jm}
\end{figure}
In Figures \ref{fig:rmse-smallad} and \ref{fig:rmse-jm}, we show the R-RMSE results for both the Small-Adult and Jm datasets, respectively. We observe that both DAWA and HDMM suffer as the number of dimensions/domain size increases because their optimizations fail to find the optimal strategy for answering the workload. In contrast, \texttt{RIPOST} maintains better performance than all of them in most tests. Regarding Identity, the noise accumulated by each query grows as the domain size increases, which is evident in both Figure \ref{fig:rmse-smallad} and \ref{fig:rmse-jm}.

\subsection{Sensitivity analysis of hyperparameters}\label{sec:sen_ana}
In this section, we analyze the behavior of \texttt{RIPOST} by evaluating its performance across different values of the hyperparameters. For this evaluation, we used a 2D tensor created from the \textit{Adult} dataset and a random workload. In each experiment, we vary one parameter while keeping the others fixed at their default values. The performance of \texttt{RIPOST} is measured using the RMSE and the number of generated blocks.
The hyperparameters and their default values are: $\epsilon = 0.1, \quad \alpha = 0.3, \quad \beta = 0.4, \quad \gamma = 0.9.$

\begin{figure*}
    \centering
    \includegraphics[width=\linewidth]{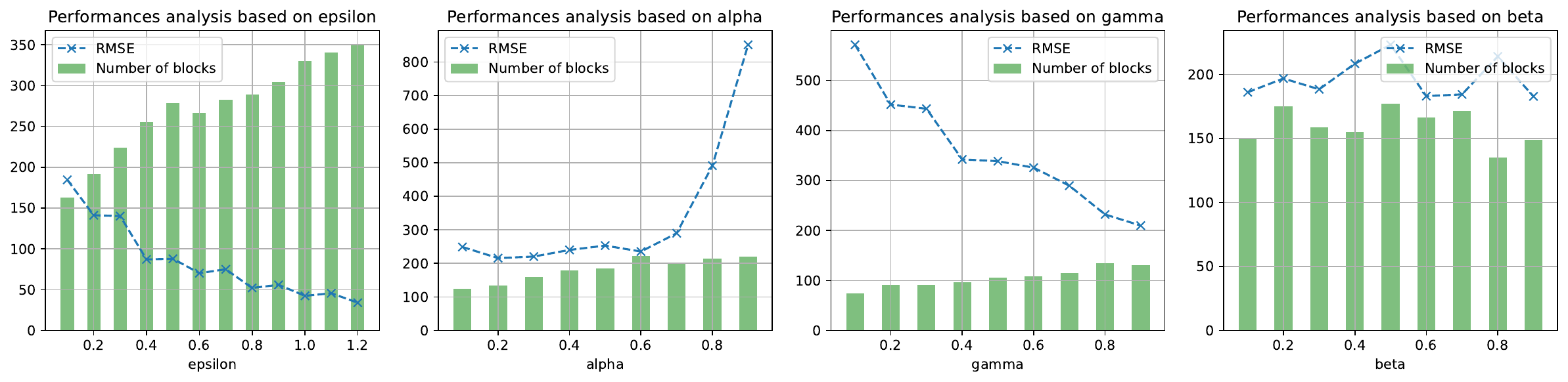}
    \caption{Analysis of hyperparameters}
    \label{fig:ana}
\end{figure*}

{\bf Privacy budget $\epsilon$:} Here, we measured the performance of \texttt{RIPOST} obtained while changing the total privacy budget allocation $\epsilon$. A smaller value of $\epsilon$ indicates stronger privacy guarantees, which results in higher noise (randomness) in the output. 

From Figure \ref{fig:ana} first plot, we observe that \texttt{RIPOST} behaves consistently with the characteristics of differential privacy. Specifically, as the privacy budget increases, the error (RMSE) decreases, and the number of generated blocks increases since $Secure\_cc$ becomes more accurate. The perturbation error (PE) did not have a significant effect in this case (more blocks generated) because, as $\epsilon$ increases, the noise added to the leaf blocks becomes smaller.

{\bf Parameter $\alpha$: } According to Algorithm \ref{alg:over}, \texttt{RIPOST} uses this parameter to distribute the budget $\epsilon$ between leaf perturbation $\epsilon_p$ and the decomposition ($\epsilon_d$). A smaller $\alpha$ means more budget is allocated to leaf perturbation and less to the decomposition, and vice-versa.  
From Figure \ref{fig:ana} (second plot), we observe that as $\alpha$ increases, more blocks are generated, which is expected. However, the RMSE also increases due to the higher perturbation error (PE). This happens because: (i) less budget is allocated to leaf perturbation, resulting in more noise, and (ii) the number of blocks increases.  
Based on the results in Figure \ref{fig:ana} (second plot), the best performance of \texttt{RIPOST} is found when $\alpha$ is between 0.2 and 0.4, achieving a balance between decomposition accuracy and perturbation error.

{\bf Parameter $\gamma$: } According to Algorithm \ref{alg:over}, \texttt{RIPOST} uses this parameter to distribute the budget $\epsilon_d$ between the first phase of the decomposition $\epsilon_1$ and the second phase $\epsilon_2$. A smaller $\gamma$ means less budget is allocated to phase 1 and more to phase 2, and vice-versa.  

From Figure \ref{fig:ana} (third plot), we observe that as $\gamma$ increases, more blocks are generated, and there is a decrease in RMSE. This highlights the importance of the first phase of the decomposition in \texttt{RIPOST}, which aims to separate the non-empty cells from the empty ones. This phase is one of the most distinctive features of \texttt{RIPOST} compared to other existing decomposition algorithms, and these results confirm the observations made in Section \ref{sec:ps}.

\textbf{Parameter $\beta$:} According to Algorithm \ref{alg:over}, \texttt{RIPOST} uses this parameter to allocate the decomposition budget for each phase, $\epsilon_1$ (or $\epsilon_2$), between $Secure_{cc}$, $\epsilon_{cc}^1$ (or $\epsilon_{cc}^2$), and $Secure_{ss}$, $\epsilon_{ss}^1$ (or $\epsilon_{ss}^2$). A smaller $\beta$ means less budget is allocated to $Secure_{cc}$ and more to $Secure_{ss}$, and vice versa.

From Figure \ref{fig:ana} (fourth plot), we observe that the performance of \texttt{RIPOST} in terms of RMSE and the number of blocks does not follow a clear trend, and fluctuates for each value of $\beta$. This suggests that both $Secure_{cc}$ and $Secure_{ss}$ play equally important roles in achieving optimal results.

\section{Discussion}\label{sec:dis}
In this section, we put forward some discussion points that can be considered as possible extensions and improvements that can be added to our proposed solution. First, according to the budget distribution strategy introduced in Section \ref{sec:bd}, both $Secure_{cc}$ and $Secure_{ss}$ will never consume the full budget allocated to them. This is due to the fact that the decomposition eventually converges; even if the depth $h$ varies from tensor to tensor, it can never be infinite. A simple optimization would be to take the unconsumed budget during the decomposition and add it to $\epsilon_p$ to reduce the noise added to the leaves, thus reducing the perturbation error (PE). On the same track of improving budget consumption, an improvement could be to apply $Secure_{cc}$ normally once and skip it (or give it budget $0$) for a certain number of steps $k$, and afterwards apply it normally once again, and so on. This will allow us to chop the budget for $Secure_{cc}$ in both phases by $k$ and add the saved budget to $\epsilon_p$ to reduce the PE. 

\begin{figure}
    \centering
    \includegraphics[width=0.5\linewidth]{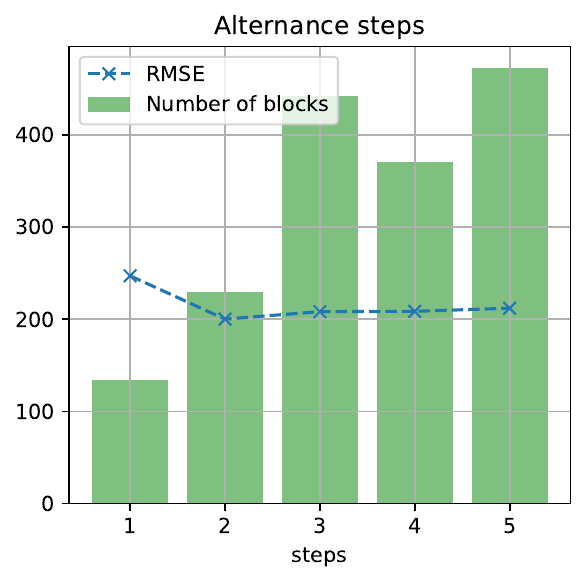}
    \caption{Results of extended \texttt{RIPOST}}
    \label{fig:steps}
\end{figure}
In Figure \ref{fig:steps}, we present the results of a small experiment conducted on a 2D tensor created using the Adult dataset. On the x-axis, we display the number of steps $k$ between consecutive applications of the $Secure_{cc}$. From the results, we observe the following: (i) as $k$ increases, the number of blocks generated also increases, which leads to a reduction in the aggregation error (AE), and (ii) the RMSE decreases, highlighting the advantage of reallocating the budget gained to reduce the perturbation error (PE). Another interesting direction for improving both decomposition computation cost and privacy consumption is to focus on enhancing the algorithm for $Secure_{ss}$. Specifically, in Algorithm \ref{alg:s_ss}, line 3, \texttt{RIPOST} iterates over all possible cutting points in the domain. A potential improvement could involve considering a smaller search space, such as a random sample of the domain. By selecting this sample randomly, we could enforce privacy and reduce budget consumption \cite{hamid}. However, such enhancements require careful consideration, as random sampling may lead to suboptimal partitioning. For this reason, we prefer to address this in a dedicated future project. The final point we would like to mention is the potential use of the generated tree to improve both query processing time and accuracy, as highlighted in \cite{cormode, privtree}. In the current version of \texttt{RIPOST}, the tree can be leveraged to enhance processing time. However, \texttt{RIPOST} does not yet incorporate the tree to improve accuracy. Given the results obtained in Section \ref{sec:eval}, it would be interesting to explore whether we can extend \texttt{RIPOST} to also address accuracy in future work.
\section{Conclusion}\label{sec:conclusion}
In this work, we addressed the problem of releasing a privacy-preserving view of multidimensional data using a domain decomposition-based method. We introduced our solution, \texttt{RIPOST}, which overcomes the design and theoretical limitations of previously proposed solutions in the literature, specifically regarding data-aware splitting and the independence from the depth $h$ of the decomposition. Through extensive experiments, we demonstrated that \texttt{RIPOST} outperforms existing methods in terms of the utility of the generated view. The positive results obtained in this work encourage us to explore various extensions and improvements in future work, particularly focusing on tighter budget management and reducing computational costs.

% PAPER ENDS

%%
%% The next two lines define the bibliography style to be used, and
%% the bibliography file.
\bibliographystyle{ACM-Reference-Format}
\bibliography{main-edbt2026}

%%
%% If your work has an appendix, this is the place to put it.
%% Please note that all the content must fit within the page limits, including any appendices.
%\appendix
%
%\section{Research Methods}
% ...

\end{document}